\documentclass{emulateapj}  
\usepackage{epsf}
\usepackage{latexsym}  
\usepackage{epsfig}
\usepackage{natbib}
\usepackage{epstopdf}
\usepackage{color}

\newcommand{\be}{\begin{equation}}
\newcommand{\ee}{\end{equation}}
\newcommand{\bea}{\begin{eqnarray}}
\newcommand{\eea}{\end{eqnarray}}
\newcommand{\bc}{\begin{center}}
\newcommand{\ec}{\end{center}}

\newcommand{\mau}{h^{-1} M_\odot}

\newcommand{\muksq}{\mu {\rm K}^2}

\begin{document}

\shorttitle{SZ bispectrum}
\shortauthors{Bhattacharya et. al.}

\author{Suman Bhattacharya\altaffilmark{1,2}, Daisuke Nagai\altaffilmark{3,4}, Laurie Shaw\altaffilmark{3,4}, Tom Crawford\altaffilmark{5}, Gilbert P. Holder\altaffilmark{6}}
\affil{$^1$ High Energy Physics Division, Argonne National Laboratory, Argonne, IL 60439}
\affil{$^2$ Kavli Institute for Cosmological Physics, The University of Chicago, Chicago, IL 60637}
\affil{$^3$ Department of Physics, Yale University, New Haven, CT 06520} 
\affil{$^4$ Yale Center for Astronomy \& Astrophysics, Yale University, New Haven, CT 06520}
\affil{$^5$ Department of Astronomy and Astrophysics, The University of Chicago, Chicago, IL 60637}
\affil{$^6$ Department of Physics, McGill University, Montreal, QC H3A 2T8, Canada}

\title{Bispectrum of the Sunyaev-Zel'dovich Effect}

\begin{abstract}

We perform a detailed study of the bispectrum of the Sunyaev-Zel'dovich effect. Using an analytical model for the pressure profiles of the intracluster medium, we demonstrate the SZ bispectrum to be a sensitive probe of the amplitude of the matter power spectrum parameter $\sigma_8$. We find that the bispectrum amplitude scales as $B_{\rm tSZ} \propto \sigma_8^{11-12}$,  compared to that of the power spectrum, which scales as $A_{\rm tSZ} \propto \sigma_8^{7-9}$. We show that the SZ bispectrum is principally sourced by massive clusters at redshifts around $z\sim 0.4$, which have been well-studied observationally.  This is in contrast to the SZ power spectrum, which receives a significant contribution from less-well understood low-mass and high-redshift groups and clusters. Therefore, the amplitude of the bispectrum at $\ell \sim 3000$ is less sensitive to astrophysical uncertainties than the SZ power spectrum. We show that current high resolution CMB experiments should be able to detect the SZ bispectrum amplitude with high significance, in part due to the low contamination from extra-galactic foregrounds. A combination of the SZ bispectrum and the power spectrum can sharpen the measurements of thermal and kinetic SZ components and help distinguish cosmological and astrophysical information from high-resolution CMB maps.

\end{abstract}

\section{Introduction}

The Sunyaev-Zel'dovich (SZ) effect \citep{sz80} is the principal source
of CMB temperature anisotropy power at angular scales smaller than a
few arcminutes. It comprises of two components, known as the `thermal'
and `kinetic' effects. The former arises from inverse Compton
scattering of CMB photons off hot electrons pervading galaxy cluster
environments, distorting the Planckian form of the CMB spectrum. The
kinetic SZ effect is caused by the Doppler shifting of CMB photons via
Thomson scattering off clouds of electrons with a non-zero bulk
velocity (along the line-of-sight) relative to the CMB rest
frame. Recently, both the South Pole Telescope \citep[SPT, ][]{lueker10,
  shirokoff10, reichardt11} and the Atacama Cosmology Telescope
\citep[ACT, ][]{fowler10, dunkley10} have placed constraints on the
amplitudes of the kinetic and thermal SZ contributions to the
small-scale CMB power spectrum. For instance, \citet{reichardt11}
measured the thermal SZ amplitude at 150 Ghz to be $3.69 \pm 0.65 \muksq$ and
placed an upper limit on the kSZ amplitude of $< 2.8 \muksq$ (at 95\%
confidence and assuming no SZ-point source correlation).

Measurements of the thermal SZ power spectrum probe the
abundance-weighted pressure profiles of hot gas in groups and clusters
over a wide range of redshifts. This is interesting in a cosmological
context because the dependence on the abundance of clusters produces a
sensitive scaling with $\sigma_8$; i.e., $C_l\propto \sigma_8^{(7-9)}$
\citep[][denoted as S10 hereafter]{ks02, trac10, shaw10}. However,
efforts to constrain $\sigma_8$ in this way have been inhibited by a
large theoretical uncertainty on the amplitude of the tSZ signal (for
a fixed cosmological model). Precise predictions of the SZ power
spectrum have been hampered by the uncertainties in modeling the state
of the intra-cluster medium (ICM) over a broad range of mass and
redshift, and at large cluster-centric radii
\citep[][S10]{battaglia10, trac10}.  Early models and simulations
produced predictions for the amplitude of the thermal SZ power
spectrum that are discrepant with recent measurements by more than a
factor of 2 \citep{lueker10}. Recent work, using
hydrodynamical simulations \citep{battaglia10, battaglia11}, N-body
simulations+semi-analytic gas models \citep{trac10} and analytic models
\citep[S10,][]{efstathiou11, chaudhuri11}, have significantly reduced
the tension between the observed and predicted values. However the
distribution of amplitudes between different models and simulations is
still significantly larger than the measurement errors
\citep{reichardt11}, degrading the constraints that can be placed on
cosmological parameters.  Furthermore, as current experiments do not
cover a sufficiently broad frequency range to enable the tSZ and kSZ
signals to be distinguished spectrally, the kSZ contribution must
currently be modeled and subtracted from the total SZ power, providing
an additional source of systematic uncertainty.

In this work we propose the SZ bispectrum as a new, robust method for
constraining the thermal SZ amplitude. The bispectrum, or its
real-space counterpart, the 3-point function, is widely used to
measure the deviation of a signal from Gaussianity both in the CMB and
in large scale structure data.  The bispectrum in the CMB map can be
measured by using three CMB temperature values and taking the average
over the survey sky. The primordial CMB bispectrum can be used to
search for signatures of primordial non-Gaussian fluctuations
\citep[see e.g.,][and references therein]{wmap7}. The dominant
secondary anisotropy signals in the CMB temperature power spectrum
arise from lensing of the CMB by large-scale structure, the ISW effect
and the SZ effect. These signals, either alone or from their cross
correlation, give rise to a measurable bispectrum \citep{cooray00_1,
  ks00, cooray99}.  The full non-Gaussian information (primordial or
from low-redshift structure) can be obtained by Fourier transforming the 
CMB temperature field and taking the product of three temperature values 
that form a triangle, then summing over all possible triangle configurations 
in harmonic space, assuming all harmonic modes forming the triangles 
are independent.

One can also measure a skewness spectrum in which two sides of the triangle are 
summed over to measure the skewness as a function of the third side \citep{munshi09}.
The total skewness spectrum in a CMB map consists of that generated by
CMB lensing, the thermal and the kinetic SZ effects, point sources, a 
signal arising due to the cross correlation of these components, and
any primordial contribution. A significant fraction of the kinetic SZ
power spectrum is generated by Gaussian density fluctuations in the
linear regime \citep{shaw11}. Although patchy reionization may contribute 
a significant fraction of the kSZ spectrum \citep{mesinger12}, it is reasonable 
to expect that the kinetic SZ bispectrum is very small -- the overall kSZ amplitude 
is 50-75\% smaller than the tSZ. 
The SZ skewness spectrum is thus likely to be dominated by the thermal SZ effect. 
As the primary CMB power decreases rapidly at $\ell>2000$, the CMB lensing 
bispectrum should also becomes negligible at high $\ell$.  \cite{cooray00} have 
shown that the SZ skewness can be measured using the Planck data with a 
S/N$\approx$10, assuming a perfect subtraction of the primary CMB and other
foregrounds.

In this work, we focus on the SZ bispectrum, showing that it can
provide information complementary to that from the SZ power
spectrum. First, as the SZ skewness spectrum is dominated by the
thermal SZ component it does not suffer from the observed
thermal-kinetic SZ degeneracy. A combination of the SZ skewness and
power spectra could therefore help disentangle the thermal and the
kinetic signals.  We show that the data from current
small-scale CMB experiments have the potential to constrain the
thermal SZ amplitude using bispectrum measurements with $5-10$\%
accuracy. Furthermore, a tightened constraint on the kinetic SZ
amplitude can also improve our understanding of the reionization
scenario \citep{shaw11, zahn11}.

Second, as we will show, the signal is dominated by massive clusters
at intermediate redshift for which high-precision X-ray observations
exist. This is in contrast to the power spectrum where the signal
mainly comes from the lower mass and higher redshift groups and
clusters \citep[e.g.,][]{trac10}.  The theoretical uncertainty in the
SZ skewness spectrum is thus expected to be significantly smaller than
that of the SZ power spectrum. Combined measurements of the power
spectrum and the bispectrum can thus be used to distinguish the contribution to
the power spectrum from different cluster mass and redshift ranges.

The goal of this study is to derive the SZ skewness spectrum using an
analytic model for the ICM gas and analyze its astrophysics and
cosmology dependence. In Section 2 we discuss the theory of the SZ
bispectrum, and in Section 3 we describe our model for calculating the
thermal pressure of the intra-cluster medium used in the bispectrum
calculation. In Section~4 we present our results, including the
distribution of the skewness spectrum signal over mass and redshift,
the theoretical uncertainty, cosmological dependence, and the
prospects of detectability from the current data from SPT like
surveys.  Section~5 presents summary and discussions. Throughout this
work, we assume a fiducial cosmology with zero curvature and parameters: 
$\Omega_b=0.045,\Omega_m= 0.27, \sigma_8=0.8, h=0.71, n_s=0.97,$ 
\citep{wmap7} and $w_0=-1$.

\section{Theoretical framework}
\subsection{Sunyaev-Zel'dovich effect}
\label{sec:sz}
The Compton $y$-parameter for the thermal SZ effect is  
\begin{equation}
y= \frac{\Delta T}{T_{\rm CMB}} =  \left(\frac{k_B\sigma_T}{m_ec^2}\right) \int n_e(l) T_e(l) dl,
\label{eq:y}
\end{equation}
where $l$ is the distance along the line-of-sight; $n_e$ and $T_e$ are the electron 
number density and temperature, respectively.

The thermal SZ power spectrum can be calculated by summing up the square of
Fourier transform of the projected SZ profile, weighted by the number density of clusters of a given mass and redshift,
\begin{equation}
C_l= f(x_\nu)^2\int dz\,\frac{dV}{dz}\int d \ln M\, \frac{dn(M,z)}{d \ln M} {\tilde y}^2 (M,z,\ell),
\label{eq:szcl}
\end{equation}
where $V(z)$ is the comoving volume per steradian, $f(x_\nu)= x_\nu \left( \coth(x_\nu/2)-4\right)$ 
with $x_\nu=h\nu/(k_BT_{\rm CMB})$ at frequency $\nu$  and is negative (positive) below (above) 220 GHz, $T_{\rm CMB}$ is the temperature of CMB at 
$z=0$ and $n(M,z)$ is the halo mass function of \citet{tinker08}, ${\tilde y} (M,z,\ell)$ 
is the Fourier transform of the projected SZ profile of a cluster, given by
\begin{equation}
{\tilde y}(M,z,\ell)= \frac{4\pi r_s}{\ell_s^2} \left( \frac{\sigma_T}{m_ec^2} \right)\int_0^\infty dx\, x^2P_e(M,z,x)\frac{\sin(\ell x/\ell_s)}{\ell x/\ell_s},
\label{eq:yl}
\end{equation}
where $x=r/r_s$, $\ell_s= D_A(z)/r_s$, $r_s$ is the scale radius of the 3D pressure profile, 
$D_A(z)$ is the angular diameter distance at redshift $z$ and $P_e$ is the electron pressure 
profile. Note that we use the mass function fit for 400 times the {\it mean} density, whereas 
the pressure profiles are defined at the {\it virial} overdensity \citep{bryan98}. We convert the mass function definition to the {\it virial} density in Eq.~\ref{eq:szcl}. 
The cosmology dependence of the SZ power spectrum comes from the angular diameter distance 
and the growth function while the astrophysical dependence comes from the pressure profiles.

\subsection{Bispectrum Theory}
\label{sec:bispec}
The CMB temperature fluctuations $\Delta \theta({\hat {\bf n}})$ in a certain direction, 
{$\hat {\bf n}$}, can be expanded into spherical harmonics as
\be
a_{\ell m}= \int d^2 {\hat {\bf n}} \frac{\Delta T}{T} Y^*_{lm}({\hat {\bf n}}) \;.
\ee
The angular bispectrum is 
\be
B_{\ell_1 \ell_2 \ell_3}^{m_1 m_2 m_3}=\left < a_{l_1m_1}a_{l_2 m_2}a_{l_3 m_3}\right >,
\ee 
where the angle-averaged quantity in the full sky limit can be written as
\be
B(\ell_1 \ell_2 \ell_3)= \sum_{m_1m_2m_3}  \left ( \begin{array}{ccc}
 \ell_1 &\ell_2 &\ell_3\\ 
 m_1 &m_2& m_3\end{array} \right ) B_{\ell_1 \ell_2 \ell_3}^{m_1 m_2 m_3} \;,
\ee 
which has to satisfy the conditions: $m_1 +m_2 +m_3=0$, $\ell_1 + \ell_2 + \ell_3$= even, 
and $|\ell_i-\ell_j| \le \ell_k \le \ell_i+\ell_j$. Here $B(\ell_1 \ell_2 \ell_3)$ is the bispectrum 
in the full-sky limit.  In the small angle limit, the flat-sky approximation is valid. By defining 
$b(\ell_1\ell_2\ell_3)$ to be the bispectrum in the flat-sky limit, then the correspondence 
between the full-sky and the flat-sky bispectrum is \citep{hu00},
\bea
\label{eq:Bl}
B(\ell_1 \ell_2\ell_3)&\approx&\sqrt{ \frac{(2\ell_1+1)(2\ell_2+1)(2\ell_3+1)}{4\pi}}\left ( \begin{array}{ccc}
 \ell_1 &\ell_2 &\ell_3 \;\\ 
 0 &0& 0\end{array} \right )\\\nonumber
&\times&b(\ell_1\ell_2\ell_3).
\eea 

Throughout this work, we focus on the high-$\ell$ regime (to avoid confusion from primary CMB 
and its lensing), where a flat sky approximation is valid. We will refer to the flat-sky bispectrum; namely,  $b(\ell_1\ell_2\ell_3)$ 
as the bispectrum for brevity and $B(\ell_1 \ell_2 \ell_3)$ as the full-sky bispectrum. Note that, 
for $\ell>500$, the Wigner 3-j symbol can be approximated as 
\be
\left ( \begin{array}{ccc}
 \ell_1 &\ell_2 &\ell_3\\ 
 0 &0& 0\end{array} \right )\approx \sqrt {\frac{2}{\pi}}\frac{(-1)^{L/2}}{[L(L-2\ell_1)(L-2\ell_2)(L-2\ell_3)]^{1/4}}
\label{eq:wigner}
\ee
 if $L=\ell_1+\ell_2+\ell_3$ is even, and vanishes for odd $L$. The approximate expression for Eq.~\ref{eq:wigner} is valid to better 
than 1 percent for $\ell > 500$, the $\ell$ range we are interested in this study.

The thermal SZ bispectrum is the volume integral of the cube of the Fourier transform of the 
pressure profile weighted by the halo mass function,
\begin{eqnarray}\nonumber
b(\ell_1\ell_2\ell_3)&=& f(x_\nu)^3\int dz\,\frac{dV}{dz}\int d \ln M\, \frac{dn(M,z)}{d \ln M} \\
&\times&{\tilde y} (M,z,\ell_1){\tilde y} (M,z,\ell_2){\tilde y} (M,z,\ell_3).
\label{eq:tsz}
\end{eqnarray}
We sum over the two smaller values of $\ell$s  and writing the bispectrum as a function of 
the largest $\ell$, we define the skewness spectrum as
\be
\Lambda(\ell)= \sqrt {\sum_{\ell_1 \ell_2} b^2(\ell \ell_1 \ell_2)},
\label{eq:skew}
\ee 
where $\ell + \ell_1 + \ell_2$ is even and $\ell_1 \le \ell_2 \le \ell$.

In previous work, the amplitude of the thermal SZ power spectrum has been defined in terms 
of the expected amplitude at $\ell = 3000$ for a given set of cosmological parameters
\citep{lueker10, shirokoff10, reichardt11, dunkley10}. In this study, we define $A_{\rm tSZ}$ 
as the ratio of the thermal SZ power spectrum amplitude at non-fiducial cosmology to that at 
$\sigma_8=0.8$ as
\be
A_{\rm tSZ}\equiv \frac{C_{3000}(\sigma_8)}{C_{3000}(\sigma_8=0.8)}.
\label{eq:asz}
\ee
Similarly, we define the amplitude of the skewness spectrum as 
\be
 B_{\rm tSZ}\equiv \frac{\Lambda_{3000}(\sigma_8)}{\Lambda_{3000}(\sigma_8=0.8)},
\label{eq:bsz}
\ee where $\Lambda_{3000}$ is the amplitude of the skewness spectrum at $\ell=3000$. 
For the fiducial cosmology, $B_{\rm tSZ}=A_{\rm tSZ}=1$. Given
that the skewness spectrum is proportional to the cube of the pressure
profile the expected relation between $B_{\rm tSZ}$ and $A_{\rm tSZ}$
is $B_{\rm tSZ}\propto A_{\rm tSZ}^{1.5}$, assuming similar mass and redshift
distribution of the power spectrum and the bispectrum. In
Section~\ref{sec:sz_cosmo}, we study the relation between the SZ
amplitude of the power spectrum ($A_{\rm tSZ}$) and the skewness
spectrum ($B_{\rm tSZ}$) in more detail.

\begin{figure*}    
  \begin{center}
    \begin{tabular}{cc}
     \resizebox{3.5in}{3.5in}{\includegraphics{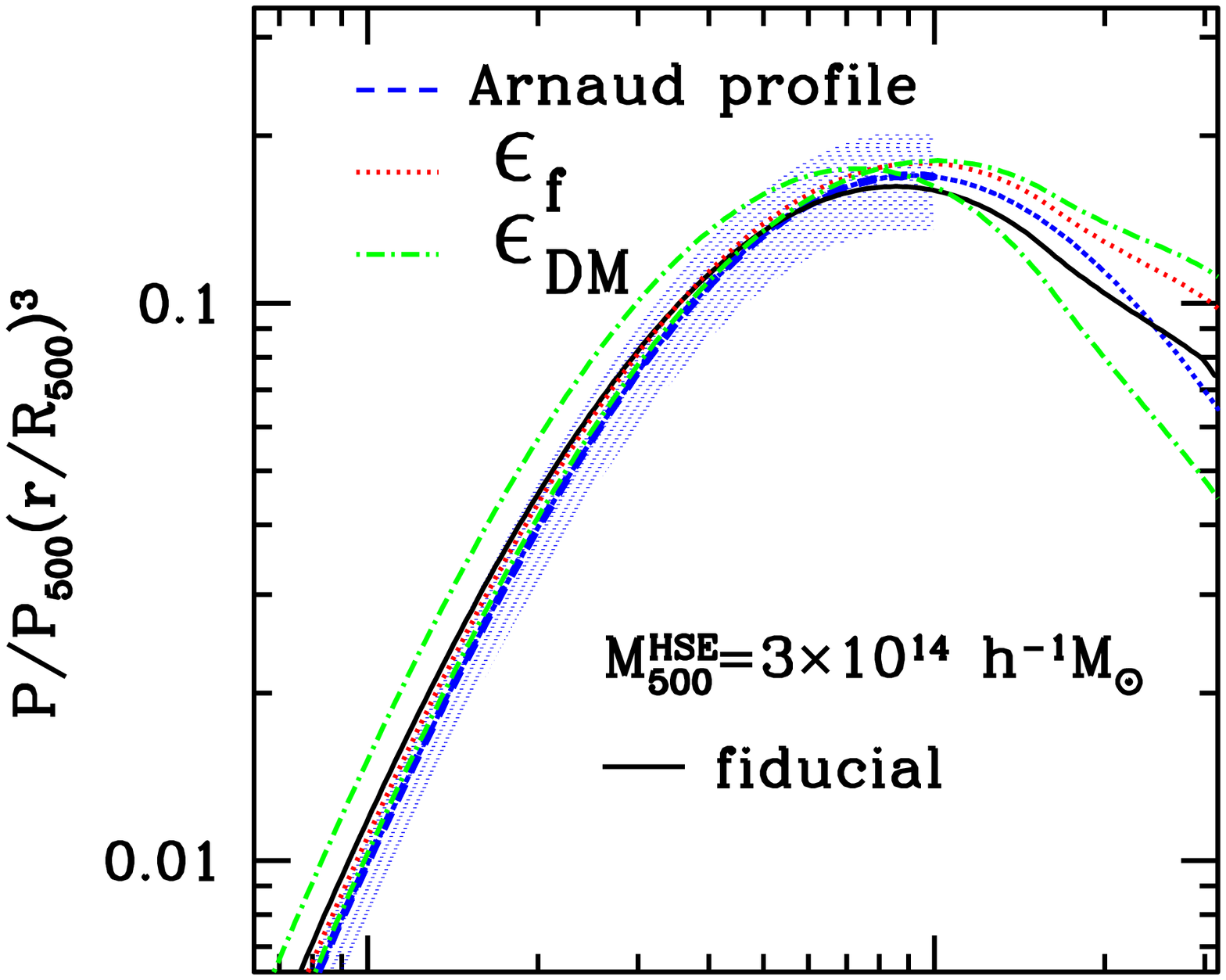}}
     \hspace{-1in} \vspace{-0.95in}
      \resizebox{3.5in}{3.5in}{\includegraphics{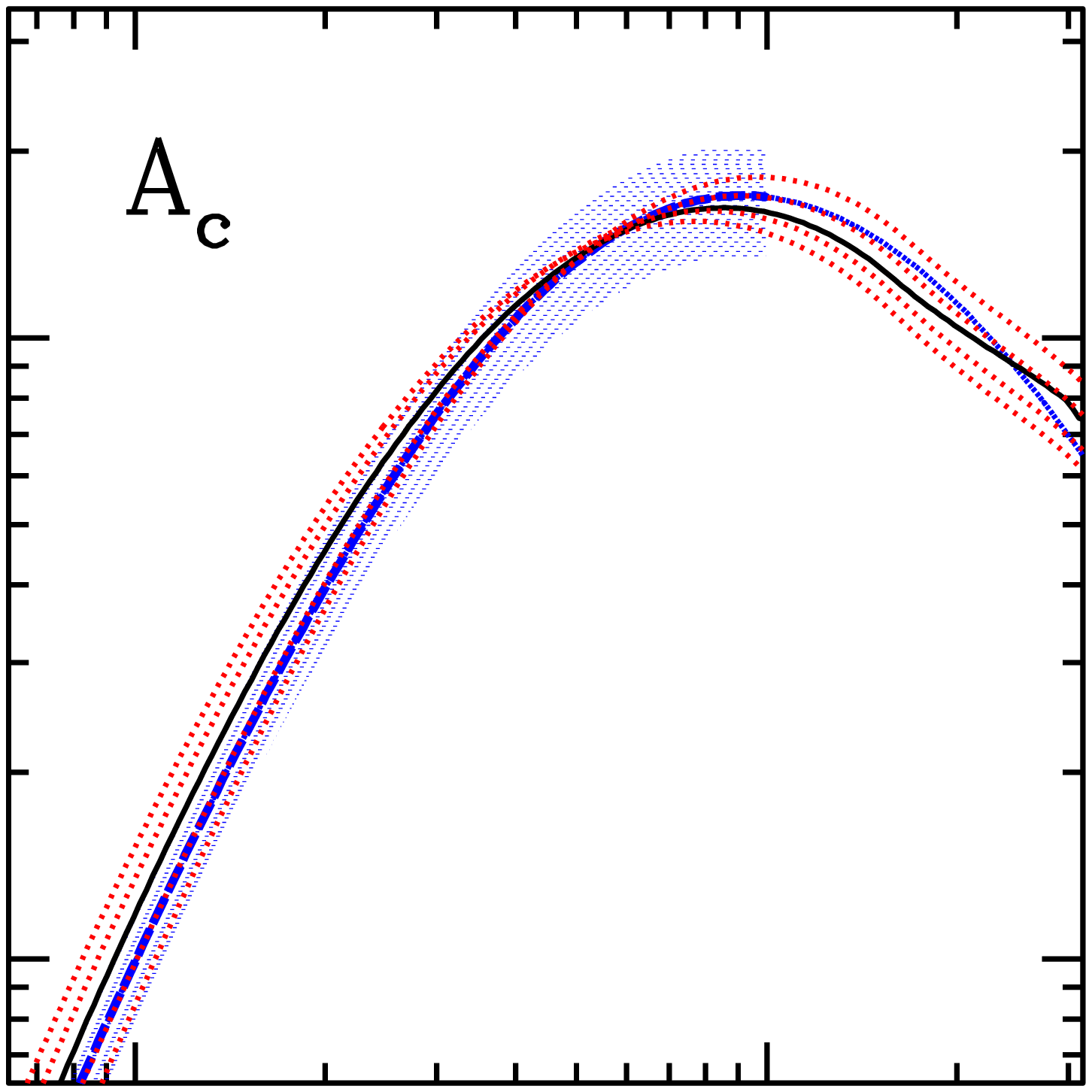}}\\
       \resizebox{3.5in}{3.5in}{\includegraphics{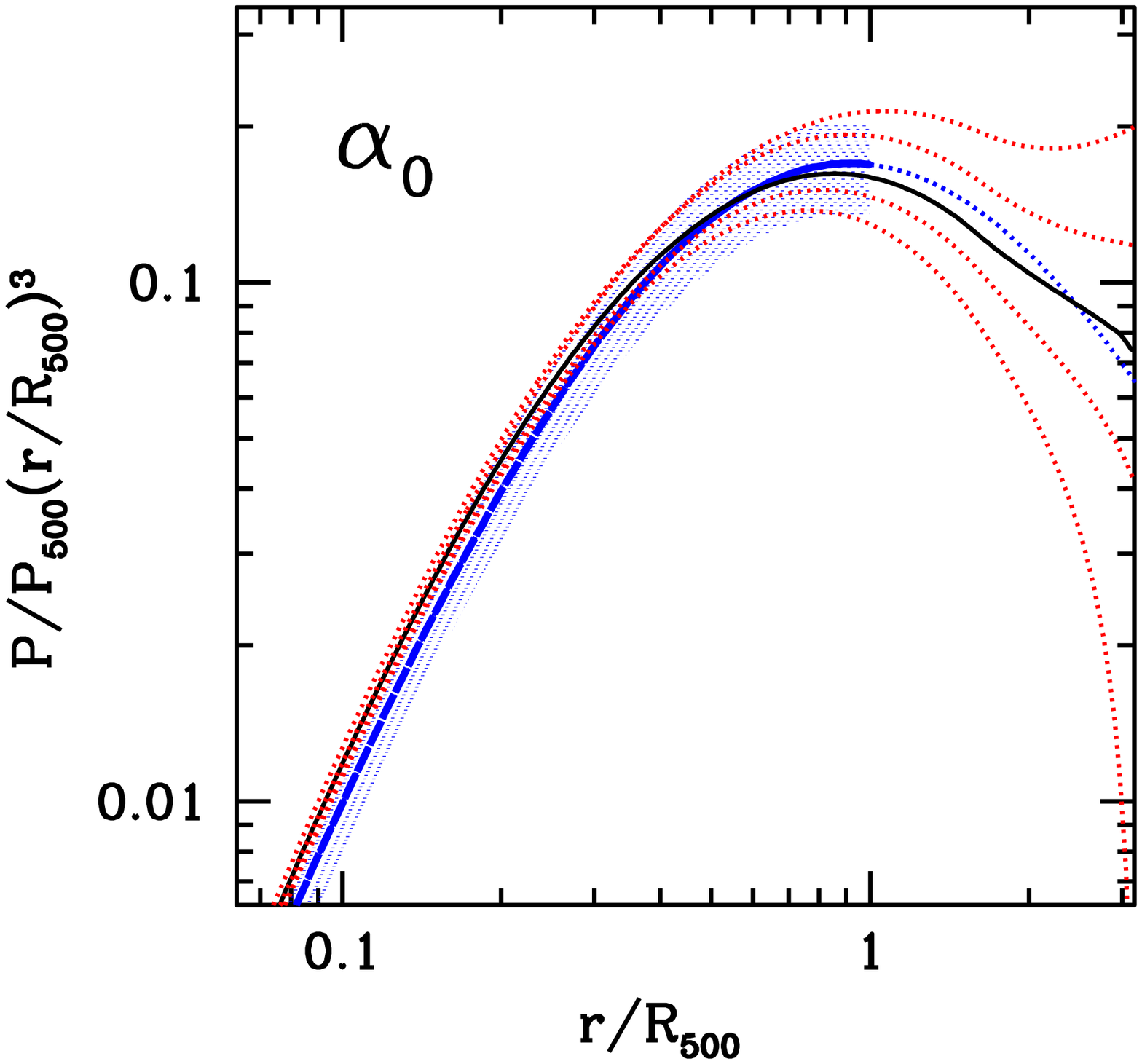}}
       \hspace{-1in}
      \resizebox{3.5in}{3.5in}{\includegraphics{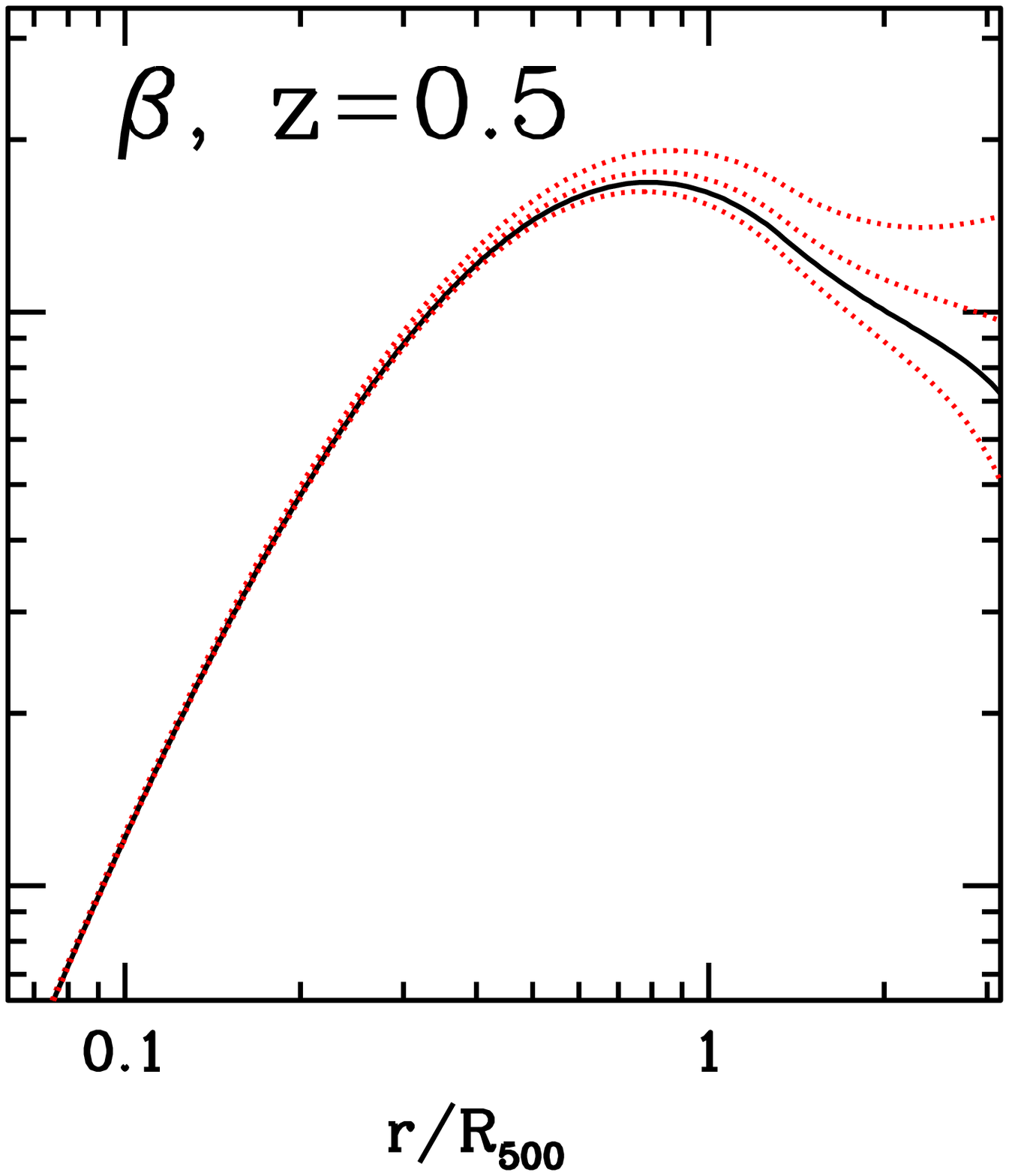}}
\end{tabular}
 \vspace{-0.2in}
\caption{Comparison between the observed `universal' pressure profile (scaled with respect to $P_{500}$) 
  of \citep{arnaud10} (blue dashed line, shaded region denotes 20\%
  scatter) and our model predictions at z=0 for a cluster of hydrostatic mass  
  $M^{HSE}_{500}= 3\times 10^{14} \mau$. In each panel we vary one
  of the input model parameters to demonstrate their effect on the
  pressure profile. The black solid line represent the
  fiducial values as given in Table~\ref{tab:param}.  The top-left
  panel shows the profiles obtained when $\epsilon_f$ is $10^{-5}$ 
  (red dotted line) and $\epsilon_{DM}$ varies from 0.2 to
  0.0 (green dashed lines, from top to bottom at $2R_{500}$); in the
  top right panel we vary $A_c$ from $0.8$, $0.9$, $1.1$, $1.2$ (red
  dotted lines, top to bottom at $2R_{500}$), the bottom-left panel
  shows the variation with $\alpha_0$ from $0.3$, $0.24$, $0.14$,
  $0.0$ (red dotted lines, bottom to top at $2R_{500}$) and the
  bottom-right panel shows the variation with $\beta$ at $z=0.5$ over
  $\{1,0.5, 0, -1\}$ where self-similar evolution is factored out (red
  dotted lines, from bottom to top). Note that the observed result is
  at $<R_{500}$, the blue dotted line is an extension of the Arnaud
  profile based on the simulation result.}
\label{fig:arnaud}
  \end{center}
\end{figure*}

\begin{table}
\begin{center}
\begin{tabular}{|l|c|c|}
\hline\hline
Parameters &  Fiducial & Range \\
\hline\hline
Dark Matter concentration ($A_C$) & $1$ & $0.8-1.2$\\
 & & \\
Energy Feedback ($\epsilon_f$) & $4\times10^{-7}$ & (1-100)$\times 10^{-7}$\\
 & & \\
Dynamical Friction Heating ($\epsilon_{DM}$) & $0.05$ & $0.0 - 0.1$\\
 & & \\
Non-thermal pressure normalization ($\alpha_0$) & $0.2$ & $0.0 - 0.3$ \\
 &  & \\
Non-thermal pressure evolution ($\beta$)& $0.5$ & $-1 - +1$\\
\hline\hline
\end{tabular}
\end{center}
 \vspace{-0.2in}
\caption{Fiducial values and the uncertainty range in the ICM parameters.} 
\label{tab:param}
\end{table}

\section{Modeling the Intra-cluster Medium}
\label{sec:icm}
To calculate the thermal pressure profiles of groups and clusters, we
use the physically motivated analytic model presented in S10. Here 
we briefly describe the key features of the model, and refer the readers 
to S10 for more details.

The model assumes that intra-cluster gas resides in hydrostatic equilibrium within the gravitational 
potential of the host dark matter halo, 
\begin{equation}
\frac{dP_{\rm tot}(r)}{dr}= -\rho_g(r) \frac{d\Phi(r)}{dr},
\label{eq:hse}
\end{equation}
where $\rho_g(r)$ is the gas density at radius $r$ from the cluster center, the total pressure, 
$P_{\rm tot}(r)=P_{\rm th}(r)+P_{\rm nt}(r)$, is given by a sum of thermal and non-thermal 
pressures, and $\phi(r)$ is the gravitational potential.  The gas is assumed to have a polytropic 
equation of state, $P_{\rm tot}= P_0(\rho_g/\rho_0)^\Gamma$, where $\Gamma=1.2$ and $P_0$ 
and $\rho_0$ are the central gas pressure and density respectively. The total mass distribution 
is assumed to follow the Navarro-Frenk-White (NFW) profile \citep{nfw96, nfw97},
\begin{equation}
\rho_{\rm tot}(r)= \frac{\rho_s}{x(1+x)^2},
\label{eq:nfw}
\end{equation}
where $\rho_s$ is the normalization constant, $x=r/r_s$, and $r_s$ is the characteristic radius 
and can be defined in terms of the concentration of halos, $r_s= r_{\rm vir}/c_{\rm vir}$ where 
``vir'' refers to the virial overdensity \citep{bryan98} with respect to the {\it critical density} of the universe. 
We adopt the concentration-mass relation given in \cite{duffy08}:
\begin{equation}
c_{\rm vir}(M,z)= 7.85A_c\left( \frac{M_{vir}}{2\times 10^{12} h^{-1} M_\odot} \right)^{-0.081} (1+z)^{-0.71},
\label{eq:conc}
\end{equation}
where $A_c$ is the normalization factor, with the fiducial value $A_c=1$. Note that the gas is 
distributed as a massless tracer of dark matter distribution. 


\begin{figure}
  \begin{center}
     \resizebox{3.2in}{3.2in}{\includegraphics{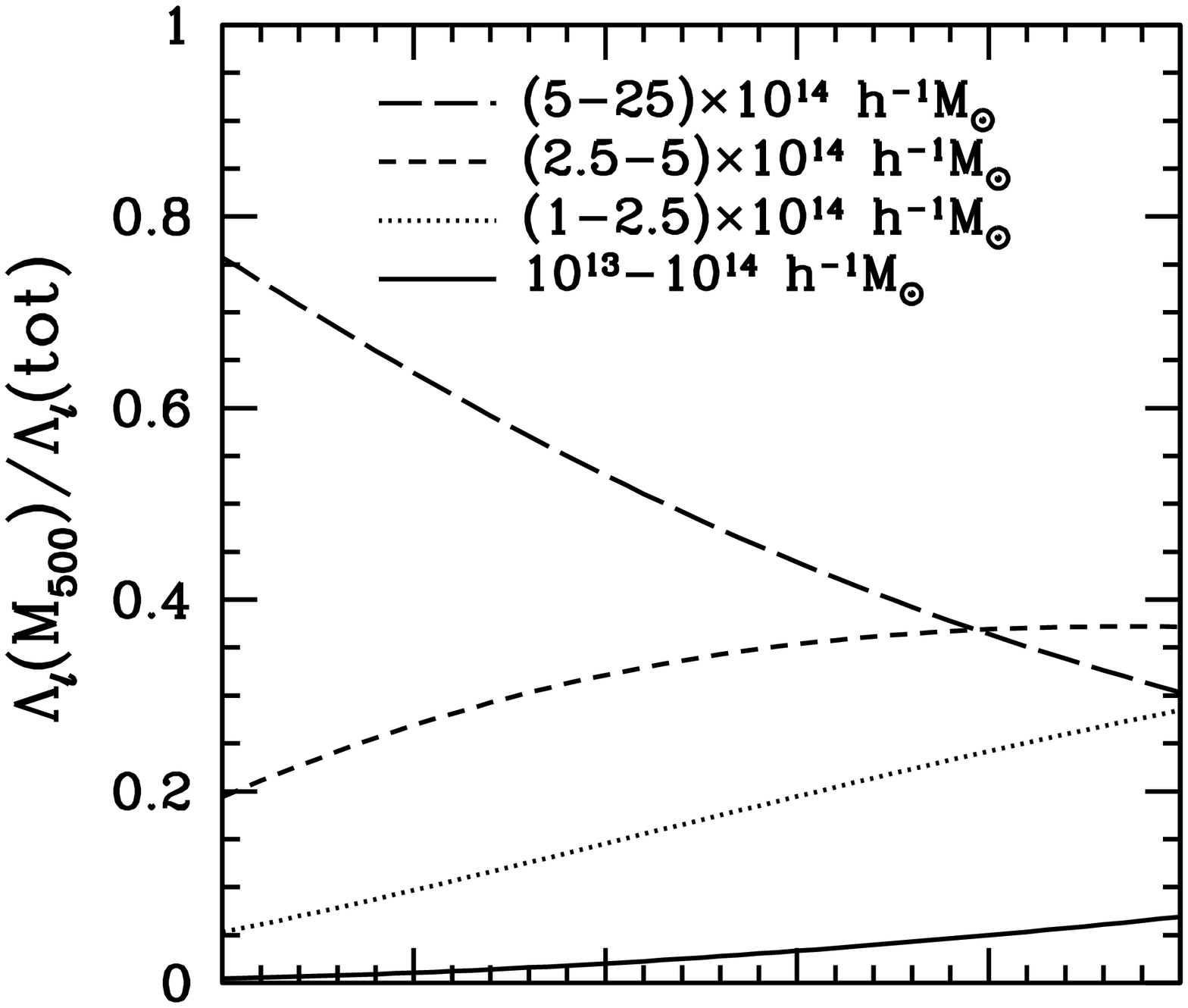}}  \\
     \vspace{-0.65in}
     \resizebox{3.2in}{3.2in}{\includegraphics{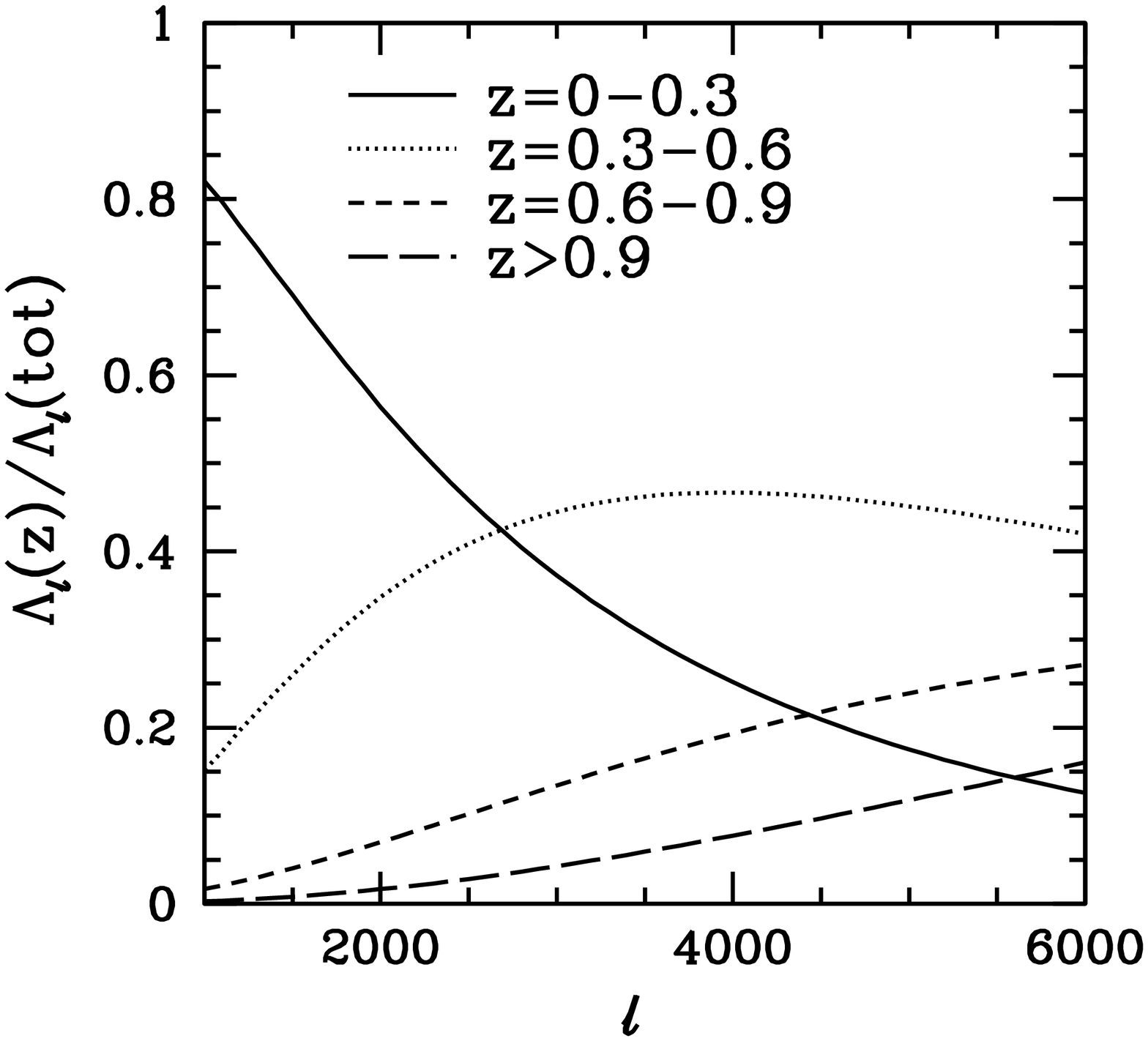}}
    \vspace{-0.2in}
    \caption{Contribution to the SZ skewness spectrum of objects within a given mass range (top) and redshift range (bottom) range. We show the ratio of the skewness spectrum in a mass/redshift range to the total skewness spectrum.}
\label{fig:szdist}
  \end{center}
\end{figure}

\begin{figure}
  \begin{center}
  \resizebox{3.2in}{3.2in}{\includegraphics{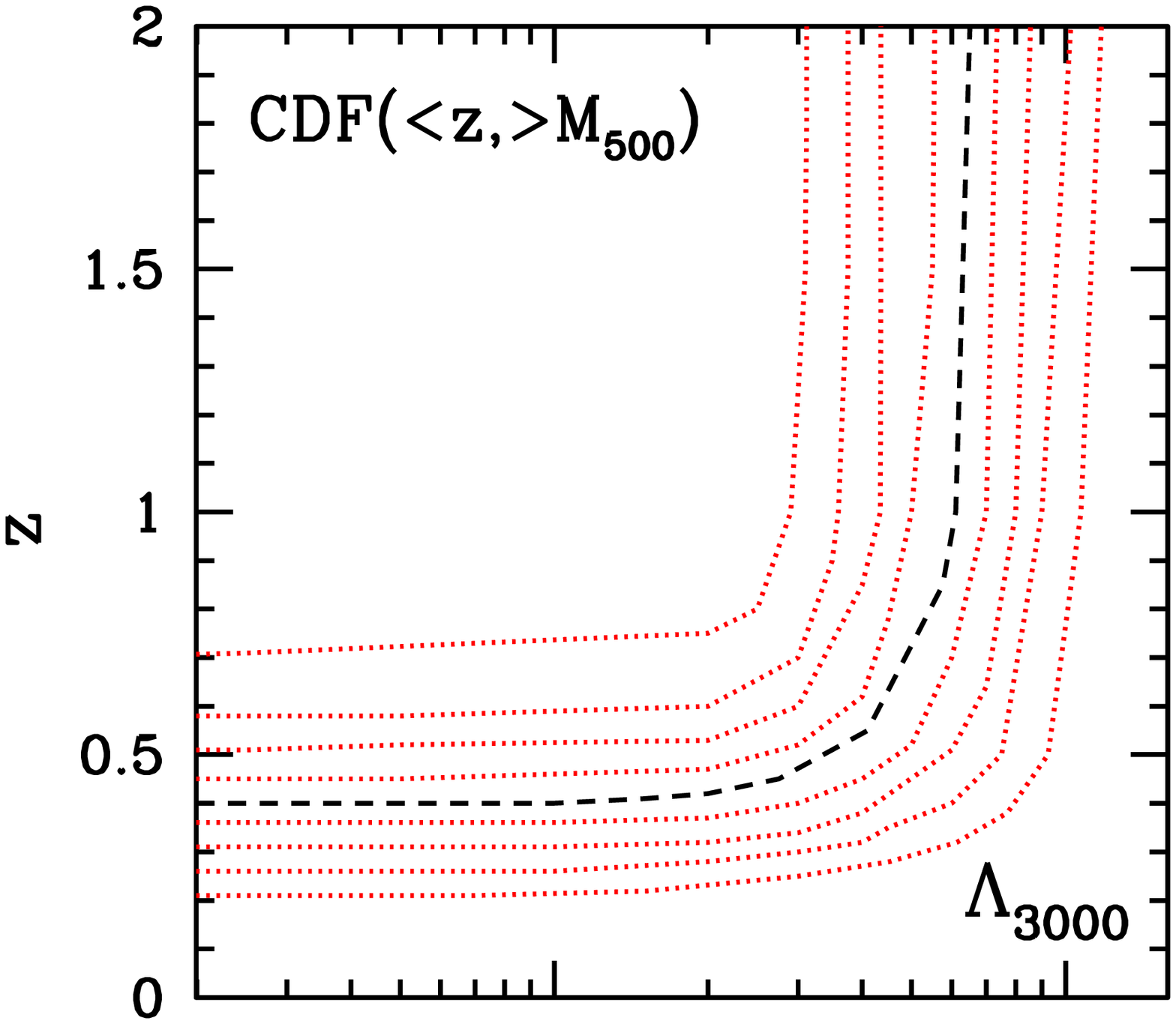}}  \\
  \vspace{-0.65in}
  \resizebox{3.2in}{3.2in}{\includegraphics{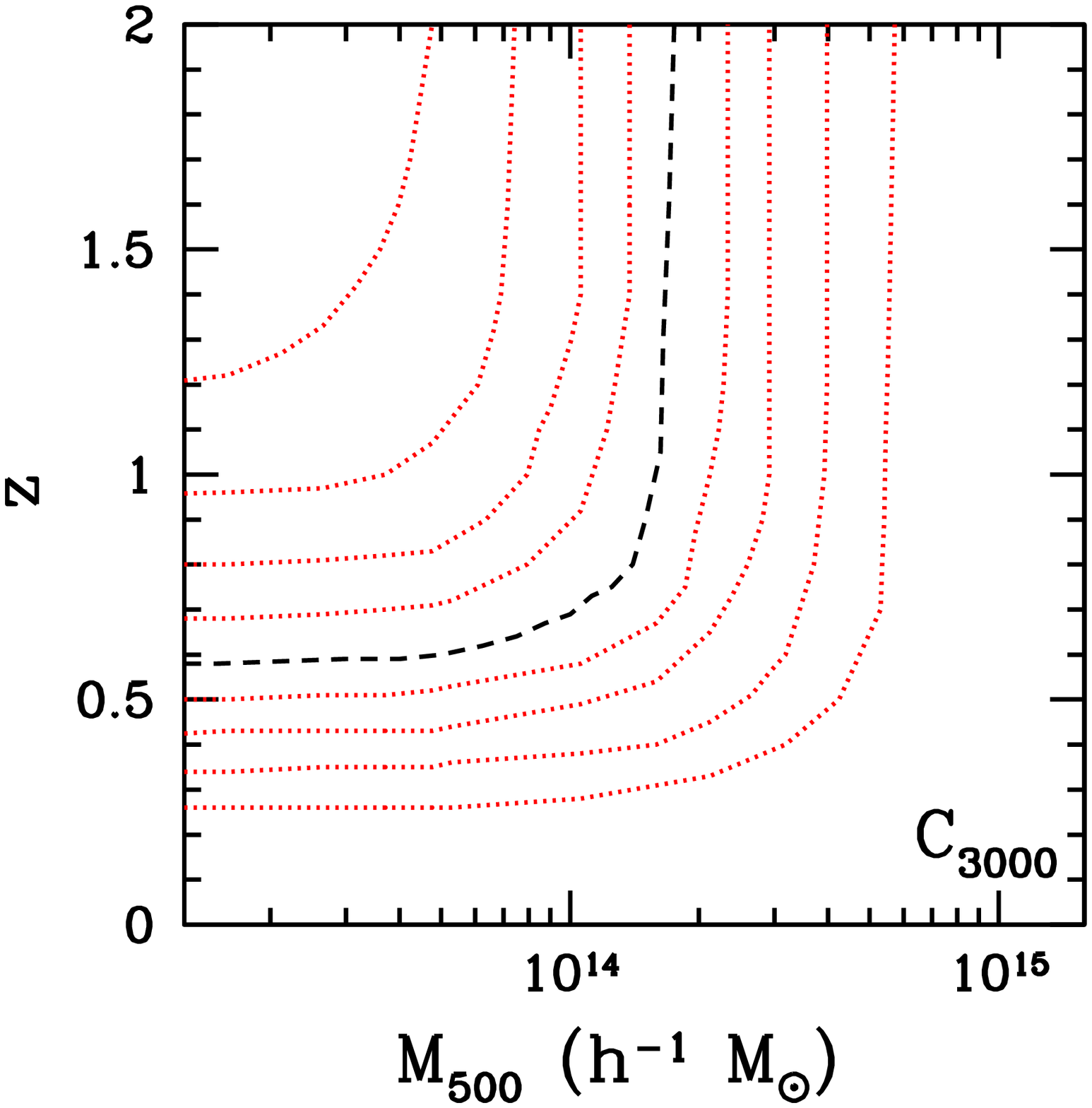}}
    \vspace{-0.2in}
    \caption{{\it Top panel:} the distribution of the skewness spectrum over mass and redshift at $\ell=3000$. At $M_{500}=5\times 10^{13} \mau$, the lines show the cumulative distribution function, CDF($>$M, $<$z)=0.1 to 0.9 from bottom to top with a step of 0.1. {\it Bottom panel}: the distribution of the SZ power spectrum at $\ell=3000$ (for comparison with the skewness distribution). The lines represent similar quantities as that in the top panel.}
\label{fig:sz_mz}
  \end{center}
\end{figure}

The model accounts for non-gravitational processes that affect the thermal properties of the ICM 
including star formation, energy feedback from supernovae (SNe) and Active Galactic Nuclei 
(AGN), and non-thermal pressure support due to bulk gas motions and turbulence. Star-formation 
is implemented by converting a certain (mass and redshift-dependent) fraction of the gas mass into stars. We adopt the observed 
stellar mass fraction for the local X-ray groups and clusters \citep{giodini09} and assume that the 
stellar populations evolve according to the ``fossil'' model of \cite{nagamine06}.
Following \citet{ostriker05}, a fraction of the rest mass
energy of the stars, $\epsilon_f M_*c^2$, is put back into the gas due
to energy feedback from SNe and AGN, where $\epsilon_f$ is a free
parameter. The model also accounts for the transfer of energy from
dark matter to gas by dynamical heating by
infalling substructures, determined by the product of the total binding energy of the halo $|E_{DM}|$ and a free parameter $\epsilon_{DM}$. Finally, we take into account the
non-thermal pressure support due to random gas motions as seen in
hydrodynamical simulations \citep[e.g.,][]{kay04, rasia04, lau09, battaglia10, nelson11}. We 
assume the non-thermal pressure to be a certain fraction of the total pressure and a power law 
in radius,
\begin{equation}
\frac{P_{\rm nt}}{P_{\rm tot}} (z)= \alpha(z)\left( \frac{r}{R_{500}} \right)^{n_{\rm nt}},
\label{eq:Pnt}
\end{equation}
where $\alpha(z)= \alpha_0f(z)$, $\alpha_0$ is the ratio of non-thermal to total pressure at $R_{500}$, 
the radius at which the spherical overdensity of the cluster is 500 times the critical density of the universe 
(enclosing mass $M_{500}$). $f(z)=\min[(1+z)^\beta, (4^{-n_{\rm nt}}/\alpha_0-1)\tanh(\beta z) +1]$, 
$n_{\rm nt}$ and $\beta$ are the radial and redshift dependence of the non-thermal pressure support, 
respectively. We set $n_{\rm nt} = 0.8$, motivated by comparisons to hydrodynamical simulations 
\citep[S10,][]{battaglia11}, leaving $\beta$ as a free parameter. 

In summary, our ICM model has five free parameters $A_C$,
$\epsilon_f$, $\epsilon_{\rm DM}$, $\alpha_0$, and $\beta$. Since the
key observable for the SZ effect is the pressure profile, we use
measurements of the pressure profiles by \cite{arnaud10} based on 31
massive ($M_{500} > 10^{14} \mau$), low redshift ($z \lesssim 0.2$)
clusters. \citet{arnaud10} showed that the pressure profiles in their
sample adheres to a universal form with $\sim$20\%
scatter. Figure~\ref{fig:arnaud} shows the pressure profile fit to
the \cite{arnaud10} data along with our predictions for a typical
cluster of mass $M^{\rm HSE}_{\rm 500} = 3\times 10^{14} \mau$ at z=0, 
where $M^{\rm HSE}_{\rm 500}$ is a cluster mass estimated assuming
the hydrostatic equilibrium (see S10 for more discussions). Each of the 5
parameters are varied over a wide range, including some extreme cases
with a very high feedback parameter or zero non-thermal pressure.

The parameter $\epsilon_f$ determines the amount of non-gravitational
(`feedback') energy injected into the ICM. High levels of feedback
inflates the gas distribution, lowering the pressure in the inner
region while boosting it in the cluster outskirts. For cluster-mass
halos, the feedback energy is a small fraction of the binding energy
and has a minimal effect on the pressure profile. The upper-left panel of Figure~\ref{fig:arnaud}
shows how $\epsilon_f$ changes the pressure profile in a cluster-mass
halo at $z = 0$. Increasing $\epsilon_f$ by factor of $\sim$ 100
changes the pressure profile by few percent at radii $R<R_{500}$ and
increases to about 30\% at $3R_{500}$. This is in qualitative
agreement with the findings from hydrodynamical simulations including
AGN feedback which indicate the impact of AGN feedback is more
pronounced in groups than in clusters \citep{quasar07,puchwein08,
  sijacki07, teyssier10,mccarthy10, battaglia10, battaglia11,
  short12}. We note however that our feedback model is highly
simplistic and does not capture the full detail of AGN feedback as
implemented in hydrodynamical simulations (for example, energy
injection via jets). 

$\epsilon_{DM}$ determines the fraction of the total dark matter energy
transferred to the ICM during mergers. Clusters frequently undergo
mergers and hence $\epsilon_{DM}$ may have a substantial effect on the
ICM physics of clusters. As shown in Figure~\ref{fig:arnaud}
(top-left), $\epsilon_{DM}$ changes the pressure profile by 20-30\%
over the entire radius range as $\epsilon_{DM}$ varies from $0-0.2$.

The normalization of the concentration-mass relation ($A_c$)
determines the characteristic radius of an NFW dark matter profile,
approximately corresponding to the radius at which the slope changes
from $-1$ to $-3$. The upper-right panel of Figure~\ref{fig:arnaud}
shows the change in the thermal pressure profile as $A_c$ varies from
$0.8-1.2$.  Increasing the concentration deepens the central potential
which draws the gas towards the cluster core, steepening the pressure
profile. The impact of increasing halo concentration provides a similar
(although opposite) effect to that of energy feedback.

The nonthermal pressure parameter, $\alpha_0$, determines the fraction
of the total gas pressure contributed by random gas motions. As
$\alpha_0$ is varied is the thermal pressure fraction adjusts
accordingly such that the total gas pressure is fixed. As shown in
Figure~\ref{fig:arnaud} (bottom-left), the thermal pressure profile
changes by about 60\% at $R_{500}$ as $\alpha_0$ varies from
$0-0.3$. $\beta$ determines the redshift evolution of the non-thermal
pressure fraction. S10 obtained an estimate for $\beta$ by comparing the
redshift 0 and 1 outputs of the simulations of \cite{lau09}, finding that the
best fit value for $\alpha_0$ increases from $0.18-0.26$ implying
$\beta=0.5$.  The bottom-right panel in Figure~\ref{fig:arnaud}
shows the pressure profile of a cluster at z=0.5 and how $\beta$
changes the thermal pressure profile.  A positive value of $\beta$
increases the amount of non-thermal pressure at higher redshift and hence
decreases the thermal pressure as redshift increases.  The pressure profile
decreases by about 20\% as $\beta$ varies from $-1$ to $1$.

Overall, we find that the range $\epsilon_f <10^{-6}$,
$\epsilon_{DM}=0.04-0.06$, $A_c=0.8-1.1$, and $\alpha_0=0.14-0.24$ to
be in agreement with the Arnaud pressure profile within the
uncertainty (see Figure~\ref{fig:arnaud}). Note that, as described in
S10, the constraints on $\epsilon_f < 10^{-6}$ and
$\epsilon_{DM}=0.04-0.06$ mainly come from the gas fraction-mass
relation of \cite{V06}. We also find the entropy scaling relation
(e.g., entropy - X-ray temperature relation) predicted by our ICM
model to be in agreement with the data of \cite{pratt10} within the
ranges of the ICM parameters.  This illustrates that the current
observations can indeed constrain the ICM parameter space
significantly.
 
\section{The SZ skewness spectrum}
\subsection{Mass and redshift contributions}
\label{sec:sz_dist}

Figure~\ref{fig:szdist} shows the fraction of skewness spectrum signal
contributed by different ranges of halo mass and redshift. At $\ell
\sim 3000$, galaxy groups ($M_{500} < 10^{14} \mau$) source only 5\%
of the signal, while lower mass clusters contribute about 20\%. The
majority of the signal comes from the massive clusters. The
distribution of the skewness spectrum is tabulated in
Table~\ref{tab:dist} over different mass and redshift ranges. We also
calculate the signal expected from clusters that are below the
detection threshold of the current generation of SZ surveys. Following
\citet{benson12, vanderlinde10}, we choose the SPT mass threshold of
$M_{500}\sim 2.5\times 10^{14} h^{-1}M_\odot$, roughly corresponding to
the detection significance of $5\sigma$.  The skewness spectrum gets only
20\% (at $\ell=1500$) to 40\% (at $\ell=6000$) of the signal from
clusters with $M_{500}\lesssim 2.5\times 10^{14} h^{-1}M_\odot$.  In
contrast, the SZ power spectrum gets about two-thirds of the signal
from groups and clusters with $M_{500}\lesssim 2.5\times 10^{14}
h^{-1}M_\odot$.

Figure~\ref{fig:sz_mz}, top panel shows the cumulative fractional distribution of the
skewness spectrum for halos with mass $<M_{500}$ and redshift $>z$, from
10-90\% of the signal (top-left to bottom-right) at $\ell=3000$. In
terms of the redshift distribution, about 90\% of the skewness
spectrum signal comes from z$<1$. The bottom panel of
Figure~\ref{fig:sz_mz} shows the distribution of the SZ power
spectrum, showing that a significant fraction of the thermal SZ power
comes from high-z, low-mass objects (i.e., about 40\% of the power
coming from $z> 0.7$ and $M_{500} < 10^{14} \mau$) where the ICM
properties are still very uncertain. In comparison, the SZ skewness spectrum signal is
dominated by low-z, massive clusters.


\begin{table}
\begin{center}
\begin{tabular}{|c|c|c|c|}
\hline\hline
Mass Range  & power & z-range & power\\
$M_{500}[\mau]$ & \% & & \%\\
\hline\hline
$<10^{14}$  & 3 & 0--0.3 & 38\\
 &  & &\\
$(1-2.5)\times 10^{14}$ &  15 & 0.3--0.6 & 45\\
 & & &\\
$(2.5-5)\times 10^{14}$ &  30 & 0.6--0.9 & 12\\
& & &\\
$(5-25)\times 10^{14}$ &  52 & $>$ 0.9 & 5\\
\hline\hline
\end{tabular}
\end{center}
 \vspace{-0.2in}
\caption{Distribution of the skewness spectrum over different mass range at $\ell \sim 3000$.} 
\label{tab:dist}
\end{table}

\begin{figure*}    
   \begin{center}
      \begin{tabular}{cc}
      \resizebox{3.5in}{3.5in}{\includegraphics{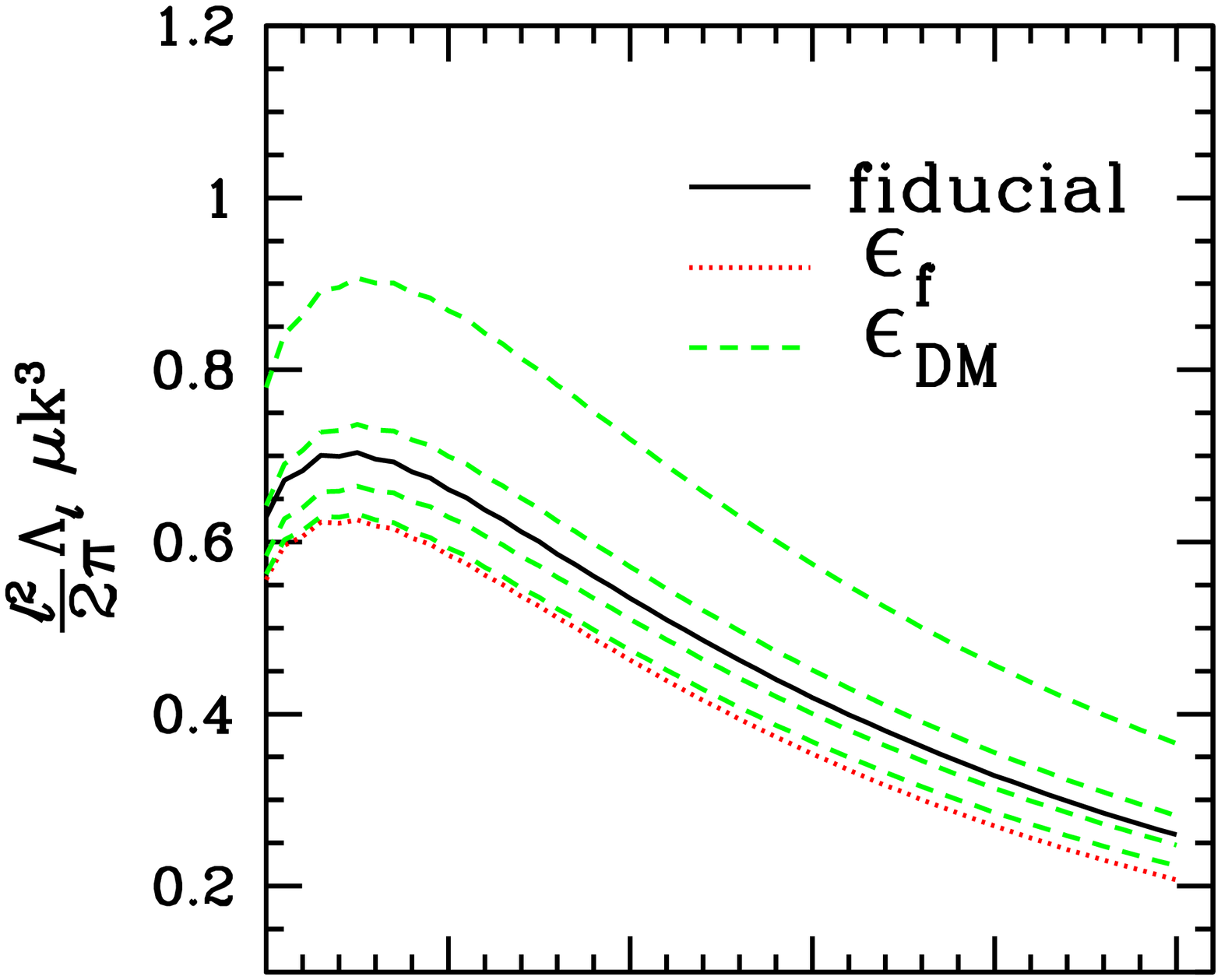}}
      \hspace{-1in} \vspace{-0.95in}
      \resizebox{3.5in}{3.5in}{\includegraphics{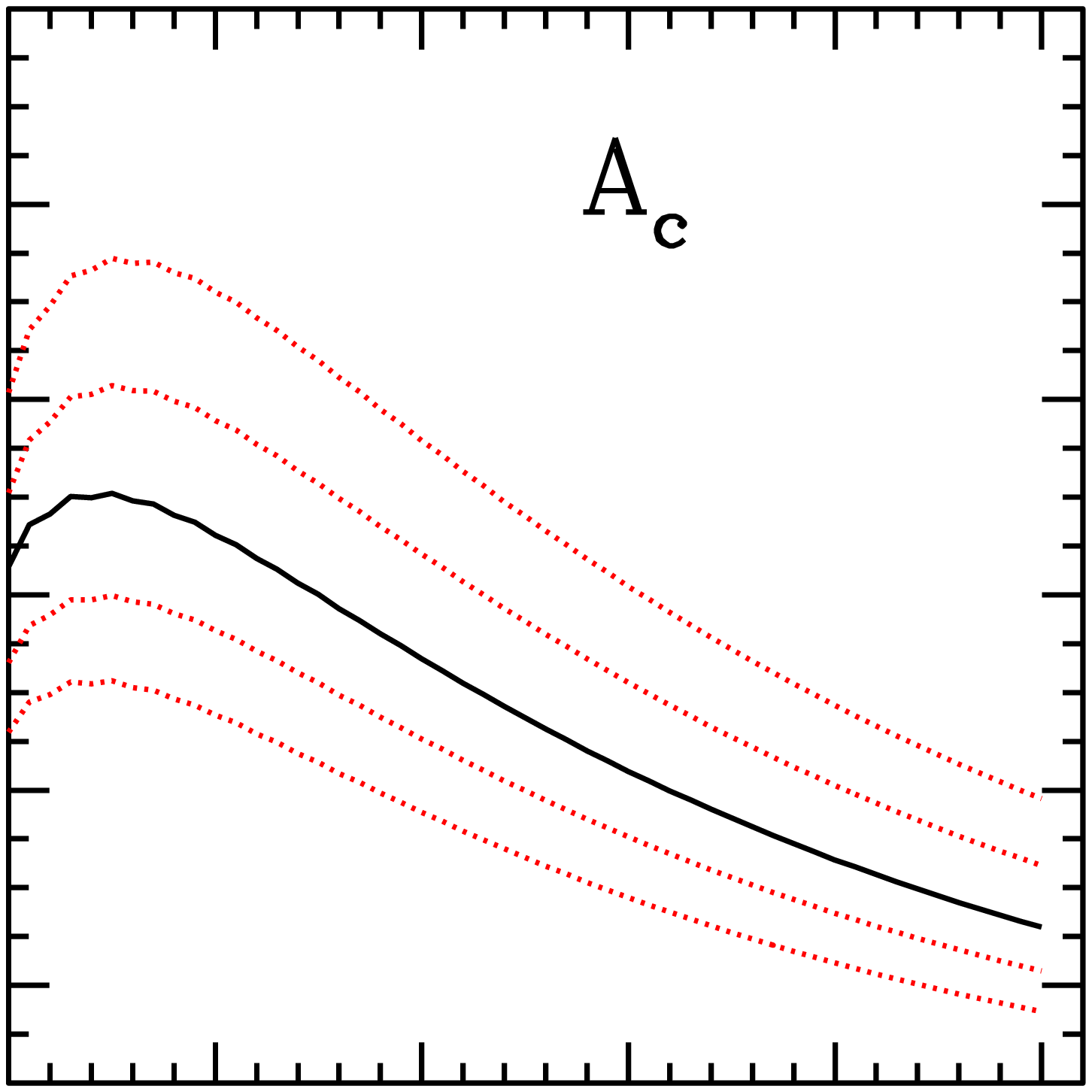}}\\
       \resizebox{3.5in}{3.5in}{\includegraphics{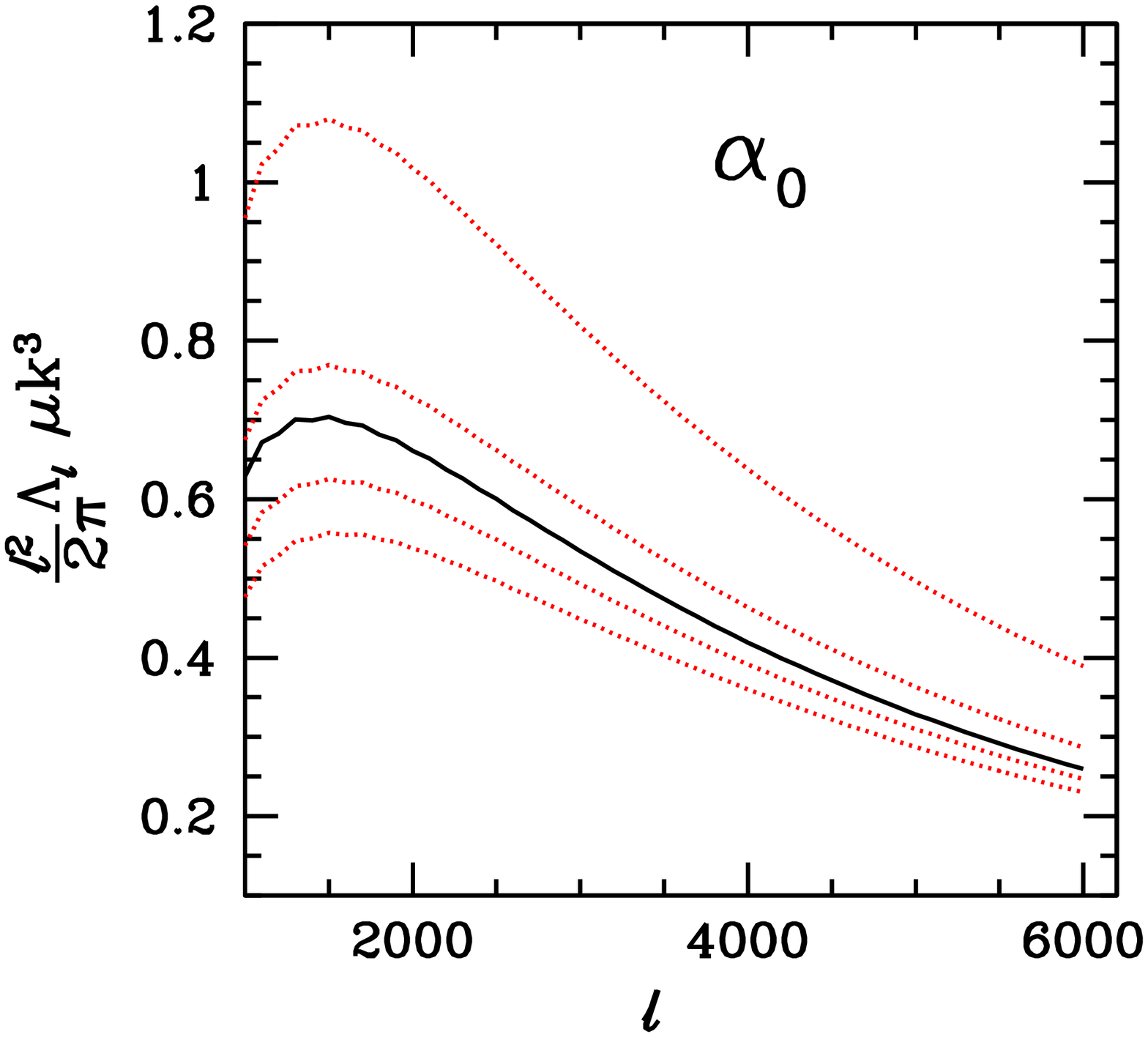}}
       \hspace{-1in}
      \resizebox{3.5in}{3.5in}{\includegraphics{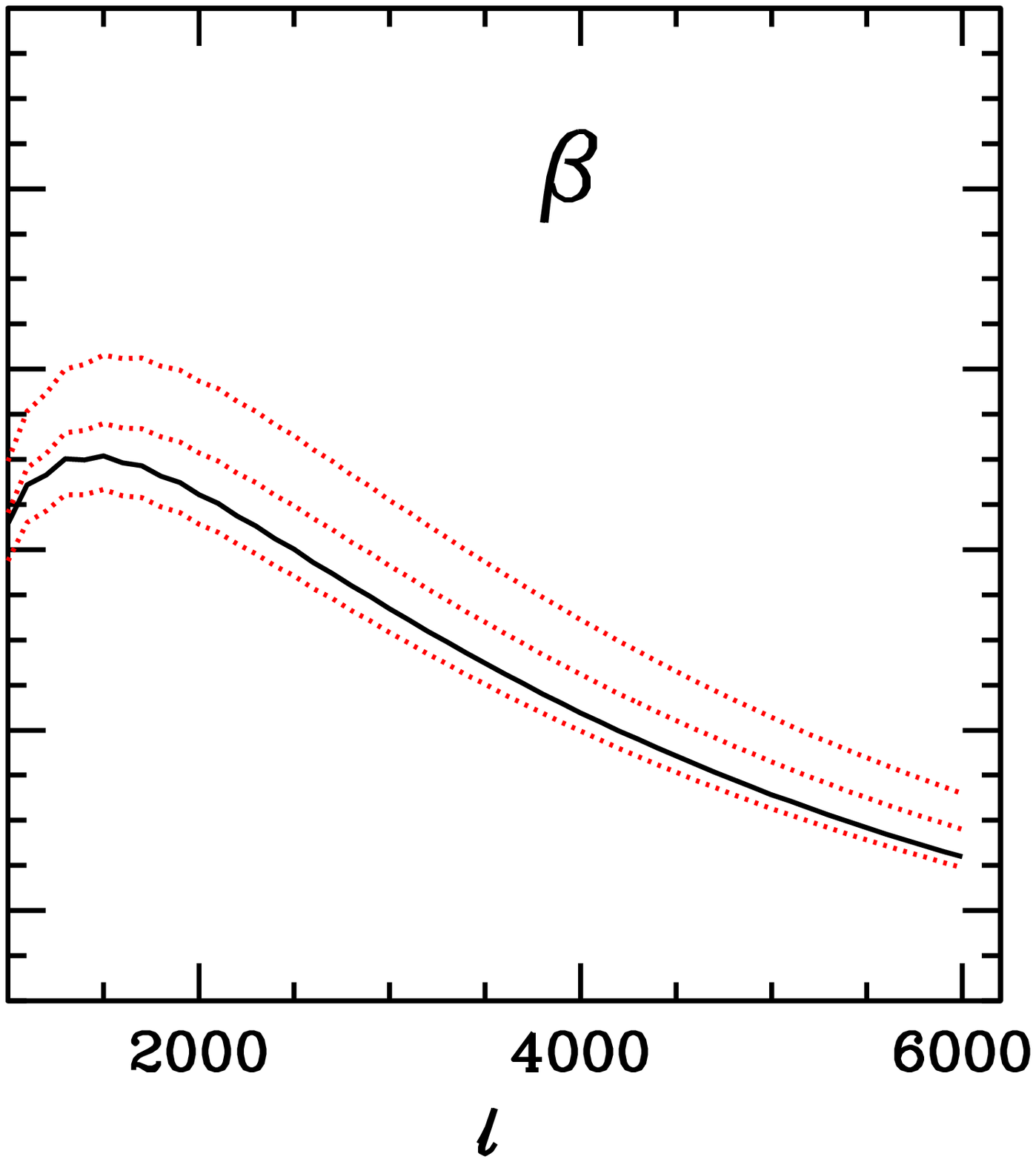}}
\end{tabular}
 \vspace{-0.2in}
\caption{Dependence of the SZ skewness spectrum on the cluster gas physics. The solid line indicates the fiducial model with $\epsilon_f=4\times 10^{-7}, \epsilon_{DM}=0.05, A_c=1.0, \alpha_0=0.2$, and $\beta=0.5$. The dotted and dashed lines represent the skewness spectrum predicted by our ICM model for a range of ICM parameter values. The top-left panel shows the impact of $\epsilon_f$ (red dotted line: $10^{-5}$ and black solid line: $4\times 10^{-7}$) and $\epsilon_{DM}$ (green dashed lines: 0.0, 0.04, 0.06, and 0.2 from top to bottom); the top-right panel shows the change with $A_c$ ($0.8, 0.9, 1.0, 1.1, 1.2$ from bottom to top); the bottom-left panel shows the variations with $\alpha_0$ ($0.0, 0.14, 0.2, 0.24, 0.3$ from top to bottom); and the bottom-right panel shows the variation with $\beta$ ($-1, 0, 0.5, 1$ from top to bottom). 
 \label{fig:blicm}}
  \end{center}
\end{figure*}

\subsection{Astrophysical Uncertainties}

We now evaluate how robust the predicted skewness spectrum is to
uncertainties in the underlying ICM model parameters.
Figure~\ref{fig:blicm} shows the variation of the SZ skewness spectrum
over the range of ICM parameters given in Table~\ref{tab:param} and
plotted in Figure \ref{fig:arnaud}. 
As discussed in Section~\ref{sec:icm}, $\epsilon_f$ does not have a
significant effect on the massive clusters that dominate the
skewness spectrum. Consequently, as shown in the top-left panel of
Figure~\ref{fig:arnaud}, the skewness spectrum varies only by 14\% as
$\epsilon_f$ varies from (1-100) $\times 10^{-7}$. 
On the other hand, the pressure profiles of massive clusters vary
substantially with $\epsilon_{DM}$, and the skewness spectrum varies
by about 40\% as $\epsilon_{DM}$ is varied from $0$ to $0.2$. In
contrast, the power spectrum receives a significant contribution from
a wider range in mass, $M_{500}=10^{13}-10^{15} \mau$, especially at
high $\ell$.  As a result, $\epsilon_f$ shows more variations at
$\ell=8000$ (group scales) than at $\ell=2000$ (cluster scales, see
Fig.~7 of S10).

Increasing the normalization of the concentration-mass relation
($A_c$) deepens the potential well of the clusters which steepens the
pressure profiles. The top-right panel of Figure~\ref{fig:blicm} shows
the change in the skewness spectrum is about $\pm$ 30\% for a $\pm$
20\% change in $A_c$ (compared to the power spectrum where the
variation is $\sim$ 20\%). This can be understood from the fact that
the skewness amplitude is $\propto$(power spectrum)$^{1.5}$, so a 20\%
change in power spectrum corresponds to a 30\% change in the skewness
spectrum.  Also note that in contrast to the case of $\epsilon_f$, the
effect of $A_c$ is not significantly mass-dependent. 

The lower-left panel shows the effect on the skewness spectrum of varying the
non-thermal pressure parameter $\alpha_0$. The skewness spectrum
changes by about 70\% across the range $\alpha_0=0-0.3$. As the SZ
skewness spectrum gets most of its signal from massive clusters at
$z \lesssim 0.4$, the uncertainty in the z-evolution of the non-thermal pressure 
($\beta$) contributes only about 20\% uncertainty to the SZ skewness spectrum. 
The SZ power spectrum, on the other hand gets its signal from high-z objects and 
as a result $\beta$ adds $\sim 30-40$\% uncertainty in the SZ power spectrum 
(bottom-right panel of Figure~7 in S10). 
 
\begin{figure}    
  \begin{center}
    \begin{tabular}{c}
      \resizebox{3.3in}{3.3in}{\includegraphics{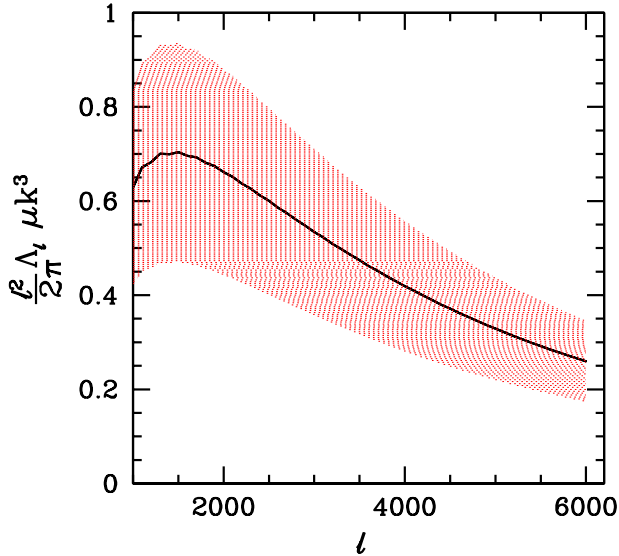}}
\end{tabular}
 \vspace{-0.2in}
\caption{Astrophysical uncertainties in the SZ skewness spectrum. The solid line represents the fiducial model and the shaded region indicates astrophysical uncertainty for the ICM parameter range calibrated by the observed pressure profile (see Section~\ref{sec:icm}). At $l=3000$, the astrophysical uncertainties are $\sim$ 33\%.} 
\label{fig:blsigma}
  \end{center}
\end{figure}

Figure~\ref{fig:blsigma} shows the astrophysical uncertainties in the shaded area around 
the fiducial model. Here, the total uncertainty is computed by adding the uncertainties 
(calibrated by the observed pressure profile and discussed in detail in Section~\ref{sec:icm}) 
in each ICM parameter in quadrature. On average, about $75-80$\% of the skewness spectrum 
signal arises from massive lower redshift objects; i.e., 
$M_{500} \gtrsim 5\times 10^{14} h^{-1}M_\odot$ at $\langle z \rangle 0.4$, and hence is less sensitive 
to the variation in ICM parameters than the power spectrum (c.f. Figure~7 in S10). 
Figure~\ref{fig:blsigma} shows the gas physics uncertainty in the skewness spectrum to be 
about 33\%. We also provide the theoretical prediction of the skewness spectrum for the fiducial 
values of cosmological and ICM parameters.

 At this point, it is worth commenting on effects of several simplifying assumptions in our model. 
First, we assumed a constant $\Gamma=1.2$. However, recent work suggests that 
$\Gamma$ could be radially dependent, especially at large radii ($r<R_{\rm vir}$). 
However, the bulk ($\approx$95\%) of the skewness signal arises from $r<R_{\rm vir}$, and 
$\Gamma \approx 1.14-1.3$ over the redshift range $z=0-1$ \citep{battaglia11}. We 
therefore find that the skewness spectrum changes by about $\pm 10-15$\% in the specified 
range of $\Gamma$.
Also, there is uncertainty in the number of baryons converted into stars in clusters, since 
this is only constrained by redshift zero observations. Secondly, the redshift dependence of the model assumes that evolution of the stellar fraction is the same for all galaxies, although it is possible that the star fraction evolve differently in a cluster environment. 
In order to assess the impact of star formation model on the skewness spectrum, we considered a  
no evolution model and found that the difference between the ``no-evolution'' model and the fiducial model (namely the ``fossil'' model of \cite{nagamine06}) is negligible. 
Finally, recent work on the thermal SZ power spectrum have shown that variations about a mean 
profile may produce up to a 15\% excess in the spectrum at $\ell \sim 3000$ \citep[e.g.,][]{battaglia11}. 
 This variation is non-Gaussian and radially-dependent and can lead to the enhancement of the bispectrum amplitude. 

\begin{figure}    
    \begin{tabular}{l}      
\resizebox{3.3in}{3.3in}{\includegraphics{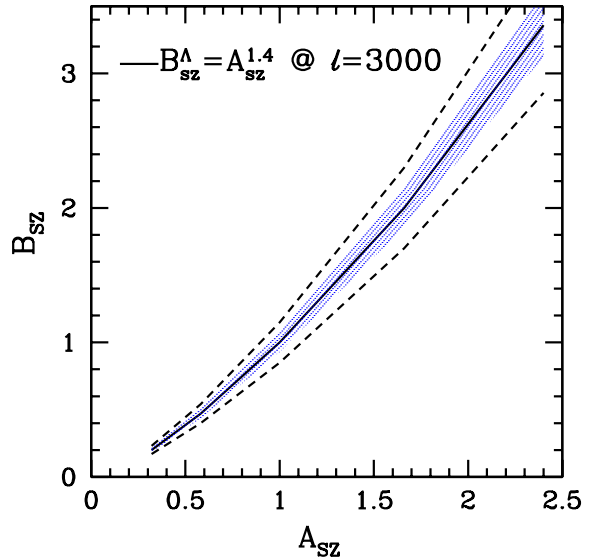}}
     \end{tabular}
 \vspace{-0.25in}
    \caption{Scaling of $B_{\rm tSZ}-A_{\rm tSZ}$ for the skewness spectrum when $\sigma_8$ varies from $0.7$ to $0.9$. The relations are relatively robust to the changes in gas physics and vary by only $\sim 7\%$ (shaded area) for the allowed ICM parameter range calibrated by the observed pressure profile (see Section~\ref{sec:icm}). The dashed lines indicate the uncertainty range bracketed by the two extreme gas physics scenarios.}
\label{fig:bszasz}
\end{figure}
\begin{figure}    
  \begin{center}
    \begin{tabular}{lll}      
\resizebox{3.0in}{3.0in}{\includegraphics{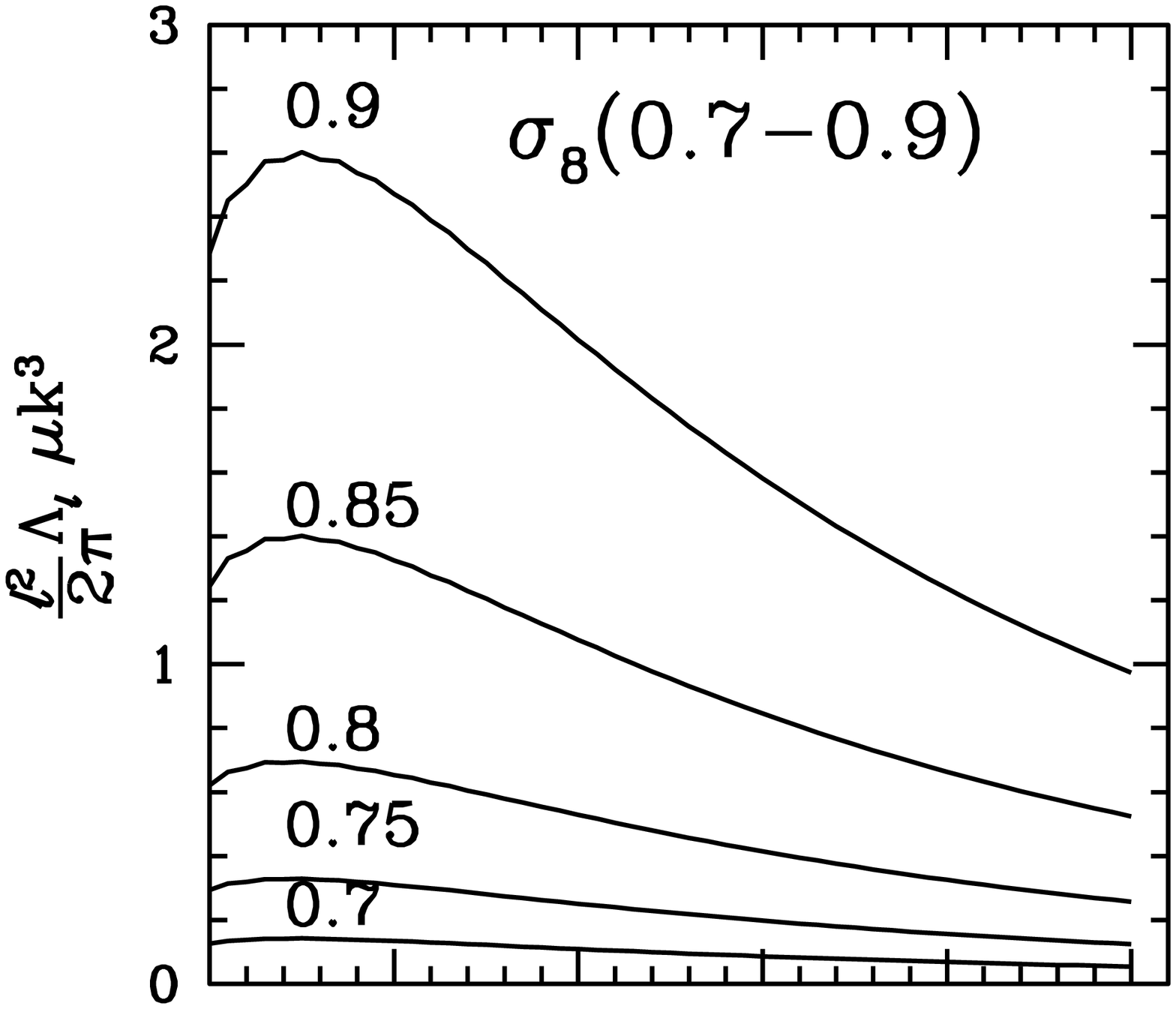}} \vspace{-0.6in} \\
\resizebox{3.0in}{3.0in}{\includegraphics{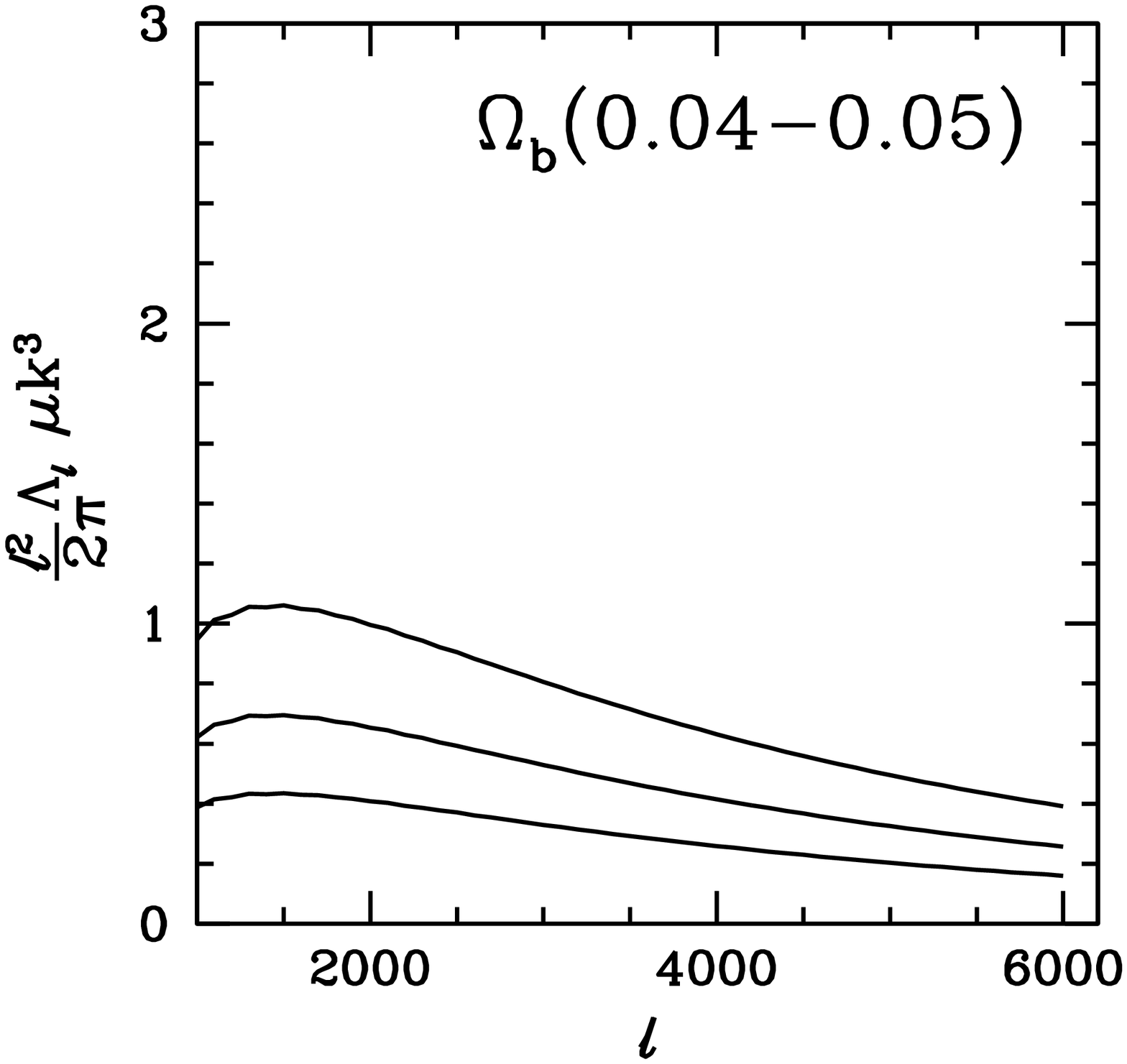}}
               \end{tabular}
 \vspace{-0.3in}
    \caption{Dependence of the skewness spectrum on $\sigma_8$ and $\Omega_b$. The fiducial values are  $\Omega_b=0.045$ and $\sigma_8$=0.8.  The range of parameters are indicated in the top-right corner of each panel. The parameter values increase from bottom to top.  
} \label{fig:bispeccosmo}
  \end{center}
\end{figure}

\subsection{The relation between the power spectrum and the skewness spectrum amplitude}
\label{sec:sz_cosmo}

\begin{table}
\begin{center}
\begin{tabular}{|c|c|}
\hline\hline
$\ell$  & $\ell^2/(2\pi)\Lambda(l) \mu {\rm K}^3$ \\
\hline\hline
1000 & 0.66\\
2000 & 0.69\\
3000 & 0.56 \\
4000 & 0.44\\
5000 & 0.35\\
6000 & 0.28\\
\hline\hline
\end{tabular}
\end{center}
 \vspace{-0.2in}
\caption{The theoretical prediction of the skewness spectrum for the fiducial values of ICM and cosmology parameters.} 
\label{tab:skewnessvalues}
\end{table}
Given a measurement of the SZ skewness spectrum amplitude, we can predict the expected amplitude 
of the thermal SZ power spectrum, namely the $B_{\rm tSZ}-A_{\rm tSZ}$ relation. Recall that the power
spectrum and the skewness spectrum are proportional to the square and the cube of the pressure profile, 
respectively. If we assume the skewness spectrum and the power spectrum have the same mass and
redshift distribution, then any changes in gas physics affects the skewness spectrum and 
(power spectrum)$^{1.5}$ in the same way. However, as shown in Sec~\ref{sec:sz_dist}, the skewness 
spectrum depends mostly on the massive clusters at intermediate redshift, while the power spectrum 
receives a substantial contribution from higher redshift galaxy groups. Furthermore, the cluster physics 
affect the gas properties in lower mass structures more than in clusters.  Therefore, the $B_{\rm tSZ}-A_{\rm tSZ}$ 
relation is expected to deviate from the above relation and depends on both cosmology and cluster astrophysics.

To investigate the dependence of the $B_{\rm tSZ}-A_{\rm tSZ}$ relation on cluster physics and 
cosmology, we vary $\sigma_8$ from $0.7-0.9$ in steps of 0.05 and compute the 
$B_{\rm tSZ}-A_{\rm tSZ}$ relation for each value of $\sigma_8$ for our fiducial gas physics model. 
Figure~\ref{fig:bszasz} shows that the skewness spectrum amplitude is proportional to the power 
spectrum amplitude,
\be
 B_{\rm tSZ}(\ell=3000)= A^{1.4}_{\rm tSZ}(\ell=3000).
\label{eq:BszAsz}
\ee
This is shallower than the expected $A^{1.5}_{\rm tSZ}$ behavior, primarily because 
the skewness spectrum signal comes from more massive clusters which are not point-like at the angular scales equivalent to $\ell \sim 3000$. We thus study 
the radial dependence of the skewness spectrum signal and find that less than 5\% signal comes from 
[1-2]$R_{\rm vir}$, 20\% comes from [0.5-1]$R_{\rm vir}$, 60\% from [0.2-0.5]$R_{\rm vir}$, the rest 15\% 
from $\le 0.2 R_{\rm vir}$. Thus a simple scaling of the pressure normalization with mass cannot explain 
the $B_{\rm tSZ}-A_{\rm tSZ}$ relation. Next we investigate the variation of the scaling relation with the 
ICM parameters. Within  the range of the ICM parameters allowed by the observed pressure profile, we 
find the $B_{\rm tSZ}-A_{\rm tSZ}$ relation is robust and the overall uncertainty does not vary by more 
than 7\% (indicated by shaded area in Figure~\ref{fig:bszasz}). The tight scaling arises because both the 
power spectrum and the skewness spectrum depend on the pressure profile. As a result, the amplitudes 
of the two spectra varies in similar ways as we vary our model parameters.

To check the robustness of the relation further, we consider two extreme gas
physics scenarios, such that the SZ power spectrum varies from $2.5-10\muksq$ 
(roughly by factor of $\sim 4$) which is roughly twice the theoretical uncertainty
assumed in the current SZ power spectrum analysis \citep{reichardt11}. 
The first case is where the pressure profile amplitude is maximally suppressed 
by high feedback and non-thermal pressure and low concentration:  
$\epsilon_f=10^{-5}$, $\epsilon_{DM}=0.1$, $\alpha_0=0.3$, and $A_c=0.8$. 
The second case other case consists of the parameter choice that predicts a 
higher amplitude: $\epsilon_f=\epsilon_{DM}=\alpha_0=0.0$ and $A_c=1.2$. 
We find that even for the two extreme gas physics scenarios the 
$A_{\rm SZ}-B_{\rm tSZ}$ has an overall uncertainty of only by $\sim 15$\% 
(indicated by dashed lines in Fig~\ref{fig:bszasz}). 
The slope of the relation changes from $1.33$ to $1.47$ between 
these two cases. This is encouraging as using the skewness spectrum data 
we can measure $B_{\rm tSZ}$, then use this theoretical relation to derive 
$A_{\rm tSZ}$. 

The relation in Eq.~\ref{eq:BszAsz} implies the skewness spectrum amplitude 
varies as $\sigma_8^{11-12}$. We also investigate how the skewness spectrum 
varies with other cosmological parameters. Figure~\ref{fig:bispeccosmo} shows 
the variation of the skewness spectrum with $\sigma_8$ and $\Omega_b$.
Each panel shows how each cosmological parameter changes the skewness 
spectrum while other parameters are fixed at their respective fiducial values. The figure 
shows that the skewness spectrum depends very sensitively on $\sigma_8$, and the 
amplitude varies as $B^{\Lambda}_{\rm SZ} \propto \sigma_8^{11.6}$, as expected. 
The variations with $\Omega_b$ is moderate, while $\Omega_m$, $h$, $w_0$, and 
$n_s$ show much smaller variation. We find the SZ amplitude for the skewness spectrum 
follows the scaling (at $\ell=3000$),
\bea
B_{\rm tSZ}&\propto&\left(\frac{\sigma_8}{0.8}\right)^{11.6}\left(\frac{\Omega_b}{0.045}\right)^{4.1}\left(\frac{h}{0.71}\right)^{2.5}\left(\frac{w_0}{-1.0}\right)^{-0.95}\\\nonumber
&\times&\left(\frac{n_s}{0.97}\right)^{-1.5}\left(\frac{\Omega_m}{0.27}\right)^{-0.46}\;.
\label{eq:scal}
\eea
Note that not all cosmological parameters follow a simple $B_{\rm tSZ} =A_{\rm tSZ}^{1.4}$ 
relation. The difference in scaling is due to the different range of masses in the mass function 
contributing to the skewness spectrum and the power spectrum.  Note that we ignore the cosmology dependence of the concentration-mass relation \citep{bh_conc11} in our study which can lead to a slightly different scaling of the skewness spectrum amplitude with cosmology. We provide the theoretical prediction of the skewness spectrum in Table~3 for our fiducial cosmology and gas parameters.

\section{Detectability}

\label{sec:detection}
\begin{figure*}    
  \begin{center}
    \begin{tabular}{cc}           
    \resizebox{3.3in}{3.3in}{\includegraphics{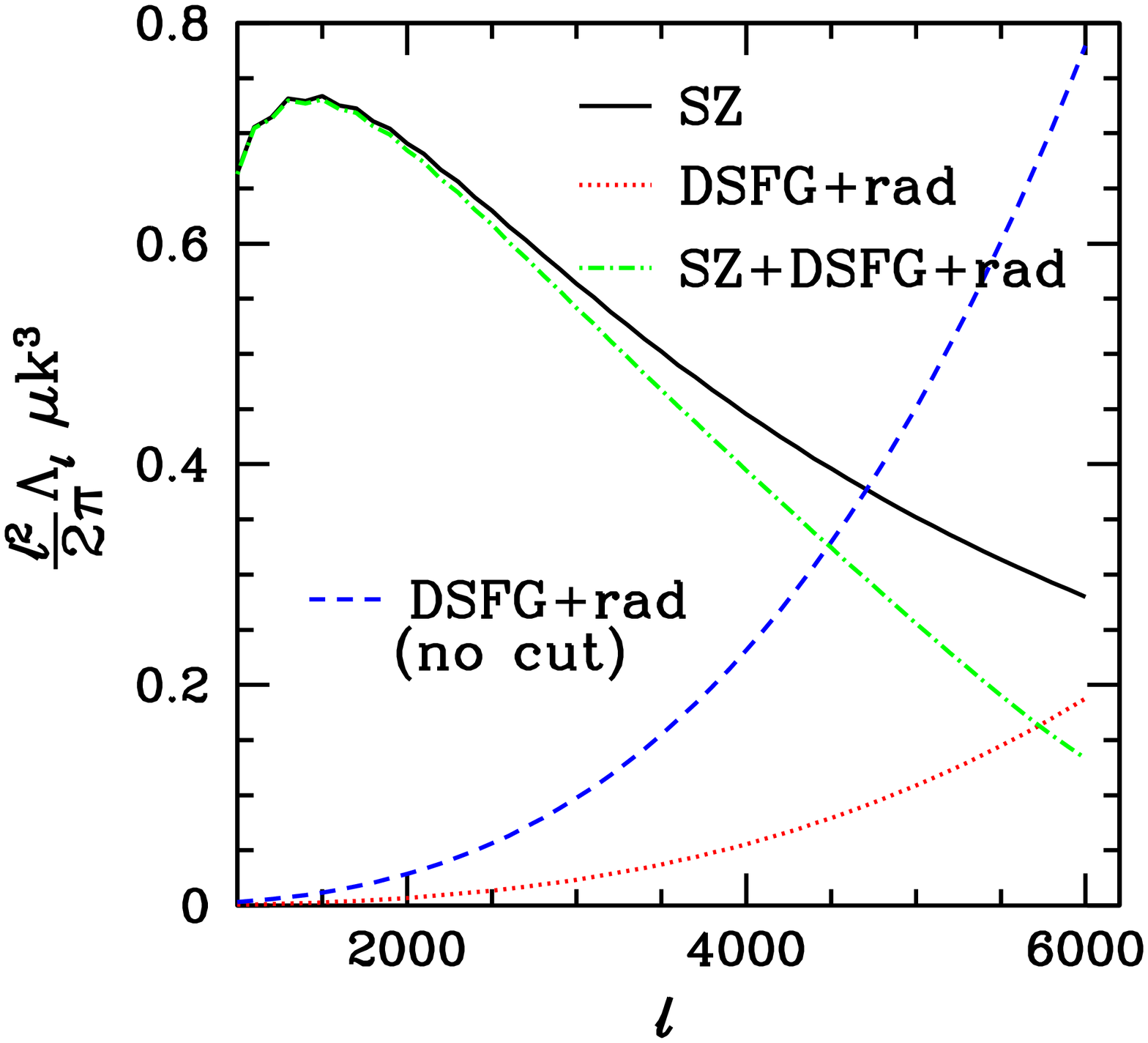}}           
    \resizebox{3.3in}{3.3in}{\includegraphics{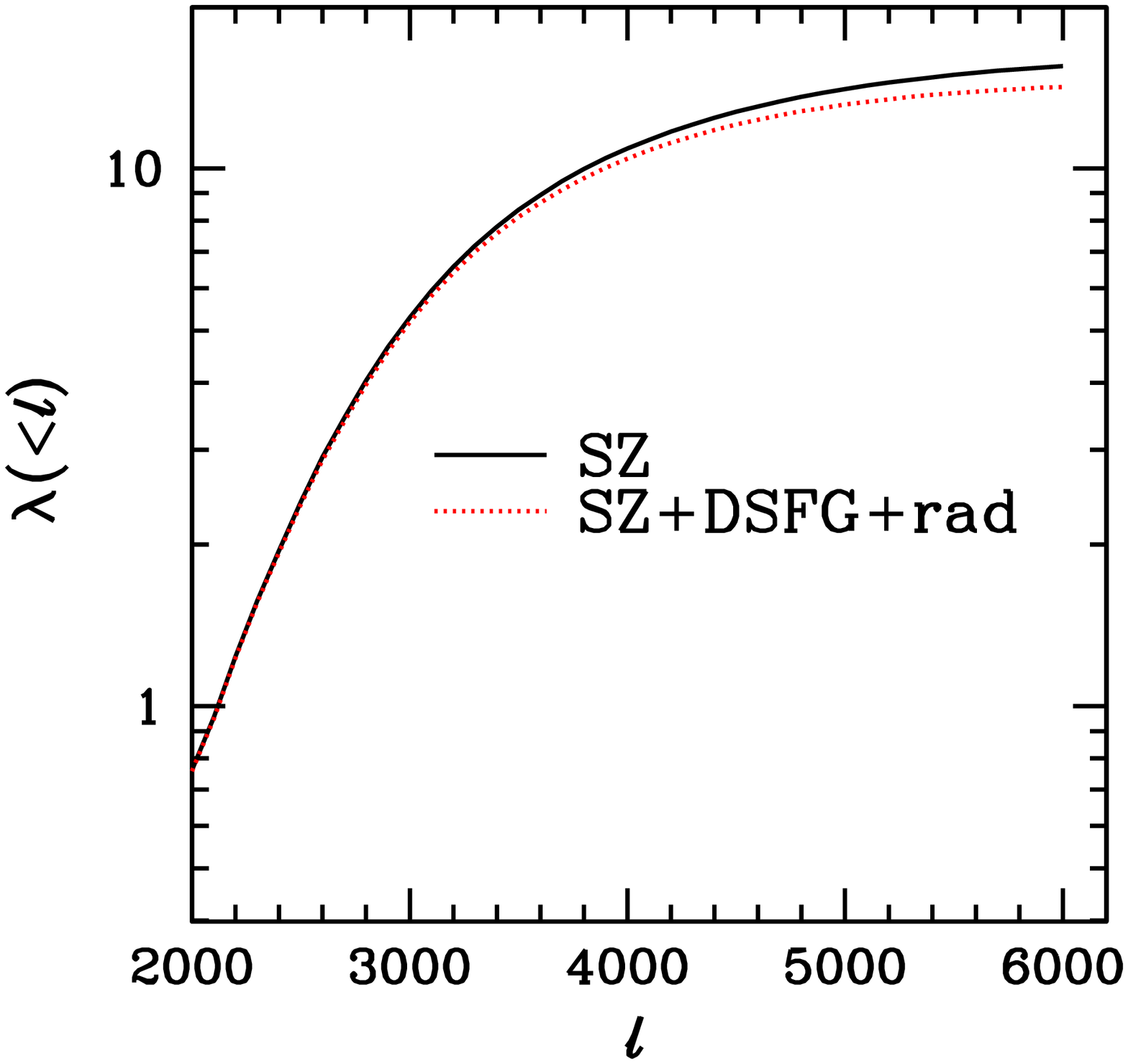}}         
    \end{tabular}
 \vspace{-0.25in}
 \caption{Detectability of the SZ skewness spectrum at 150~GHz. The left panel shows the SZ (black solid) and the point sources-DSFG+radio skewness spectrum (red dotted line). Also shown are the total skewness spectrum:  SZ+DSFG+radio (green thin dot-dashed) and the DSFG+radio skewness spectrum if no point sources are removed from the map (blue short dashed line). The right panel shows the signal-to-noise integrated to a certain $\ell$ of the SZ skewness spectrum (black solid), the total- SZ+DSFG+radio (red dotted). We assume an SPT-like survey with sky coverage of 2500~$\deg^2$, 18 $\mu$K-arcmin, 1.2 arcmin at 150~GHz.}
\label{fig:szbispec}
  \end{center}
\end{figure*}

We now assess the detectability of the SZ skewness spectrum in the
presence of other secondary CMB anisotropy signals and extra-galactic
foregrounds. Other than the thermal SZ effect, point sources (dusty, 
star-forming galaxies (DSFGs), and radio sources) will also contribute 
to the skewness spectrum signal. The main difference is while the thermal SZ effect contribute a negative bispectrum, the point sources contribute a positive signal \citep{rubino-martin03}. Assuming that the clustered component 
of the DSFGs follows the Gaussian distribution, the skewness spectrum of 
the correlated DSFG is zero.  As mentioned before, the kinetic SZ follow
approximately Gaussian distribution and is thus assumed to have
negligible skewness. We also ignore a possible skewness spectrum
signal arising from the correlation between the SZ effect and the DSFG
sources.

\subsection{Noise Estimates}

The noise of the total skewness spectrum, assuming the primary contribution to the 
variance of the bispectrum comes from the power spectrum of the CMB sky, is given by 
\citet{ks00}
\be
N^2(\ell_1 \ell_2 \ell_3)= C(\ell_1) C(\ell_2) C(\ell_3) \Delta_{\ell_1 \ell_2 \ell_3},
\label{eq:noise}
\ee
where $C(\ell)$ is the total power spectrum that includes contributions from 
(1) the SZ power spectrum computed using Eq.~\ref{eq:szcl}, (2) the lensed 
CMB power spectrum computed using CAMB for the fiducial cosmology 
\citep{camb}, (3) the noise due to the finite resolution of an experiment, and 
(4) point sources, including dusty star-forming galaxies (DSFG) and radio 
sources.  Note that $\Delta_{\ell_1 \ell_2 \ell_3}=$ 6 if all three $\ell$s are 
equal, $\Delta_{\ell_1 \ell_2 \ell_3}=$2 if two of the three $\ell$s are equal and $\Delta_{\ell_1 \ell_2 \ell_3}=$1 if all three $\ell$s are different.  We describe
the estimate of (3) and (4) in more detail below.

The noise due to the finite resolution of the experiment is given by 
\citet{knox95, jungman96}
\be
N_b(\ell)= w^{-1}\exp \left[\frac{\ell^2\theta^2}{8\log 2}\right],
\ee
where  $w^{-1}=[\sigma_{\rm pix}\theta/T_{\rm CMB}]^2$ is the weight 
per unit solid angle, $\sigma_{\rm pix}$ is the noise per pixel, $\theta$ is 
the FWHM of the instrument in radians, $T_{\rm CMB}$ is the CMB temperature. 
 Eq.~\ref{eq:noise} gives the expression for noise in the full sky limit. For the 
fractional coverage, Eq.~\ref{eq:noise} is multiplied by $f_{\rm sky}^{-3/2}$, 
where $f_{\rm sky}$ is the fraction of the sky covered by the assumed surveys.
Here we consider an SPT like survey with $f_{\rm sky}$ given by 2500~$\deg^2$ 
sky coverage, $\sigma_{\rm pix}$ is 18~$\mu$K-arcmin at 150~GHz with the 
pixel size set by a Gaussian beam of FWHM=1.2 arcmin. The powers of the unresolved 
DSFG sources at 150~GHz are given by $5.25\times 10^{-6} \muksq$ (Poisson) and 
$4.4\times 10^{-6} (\ell/3000)^{-1.2} \muksq$ (clustered), and that of the radio sources 
is $9\times 10^{-7} \muksq$ (Poisson) at 150~GHz \citep{reichardt11}. 

\subsection{Point Source Bispectrum}

The DSFG population comprises of a clustered component and a Poisson component, 
both of which contribute to the point source power spectrum. We assume that the clustered 
component produces Gaussian fluctuations and hence does not contribute to the 
bispectrum of the point sources. We thus consider the bispectrum contribution only from 
the Poisson component of the DSFGs and the radio galaxies. The Poisson component of 
the DSFG typically comprises of two populations: (1) the bright point sources which are detected 
by the CMB experiments \citep{viera09, marriage10} and (2) the faint sources which follow a 
shallower distribution compared to the bright sources, and are undetected by the current CMB 
experiments but contribute a significant portion of the arcminute scale DSFG power measured 
in the CMB experiments. Both populations are expected to contribute to the total point source 
bispectrum. 

Given the DSFG flux number counts $dN/dS$, we can write the bispectrum of DSFGs as
\be
\frac{b_{\rm DSFG}}{g^3(x_\nu)}= \int_0^{S_{\rm min}} dS\, S^3 \frac{dN}{dS}=\frac{N_{\rm DSFG}(>S_{\rm min})}{3/\beta_{\rm DSFG}-1} S_{\rm min}^3,
\label{eq:cib1}
\ee
where the flux number counts is assumed to vary as $dN/dS \propto S^{-\beta-1}$, 
$S_{\rm min}$ is the flux cut above which all sources are detected and removed from 
the CMB maps, $N_{\rm DSFG}(>S_{\rm min})$ is the number density (per unit solid angle) of 
the DSFG with $S>S_{\rm min}$, $g(x_\nu)=1/(65.55 {\rm \; MJy\; sr}^{-1})[\sinh (x_\nu/2)/x_\nu^2]^2$, 
and $x_\nu=\nu/(56.8 {\rm GHz})$. 
Similarly, we define the bispectrum of the radio sources in terms of the radio source power 
spectrum as 
\be
\frac{b_{\rm rad}}{g^3(x_\nu)}= \int_0^{S_{\rm min}} dS\, S^3 \frac{dN}{dS}=\frac{N_{\rm rad}(>S_{\rm min})}{3/\beta_{\rm rad}-1} S_{\rm min}^3.
\label{eq:rad}
\ee
We assume all point sources above $5\sigma$ at 150~GHz are detected and removed (masked). 
For the SPT 150~GHz band, this corresponds to 
$S_{\rm min}$=6.4 mJy \citep{viera09}. We adopt the number density of the DSFG to be 
$N_{\rm DSFG}= 0.3\deg^{-2}$ \citep{negrello07} and $\beta_{\rm DSFG}=1.93$ for the total 
(bright +faint) DSFG population to be consistent with the total DSFG power spectrum measured 
in the CMB surveys \citep[e.g.,][]{reichardt11}. For the radio population we adopt 
$N_{\rm rad}(>S_{\rm min})= 1.29\deg^{-2}$ and $\beta_{\rm rad}=1.03$, which is consistent with 
the 150~GHz source counts in \cite{viera09}.  The skewness spectra of the radio sources and DSFGs 
are calculated the same way as the SZ skewness spectrum by plugging Eqs.~\ref{eq:cib1} and \ref{eq:rad} 
in Eq.~\ref{eq:skew}.  

The skewness spectrum for the various components at 150~GHz are shown in the left panel of 
Figure~\ref{fig:szbispec}. Provided  the point sources that are detected above 5$\sigma$ in the SPT 
band-power are removed, the DSFG and radio population make negligible contribution to the skewness 
spectrum at $\ell \lesssim 4000$. Beyond this scale the skewness spectra of point sources 
and SZ effect become comparable. If no point sources are removed, then the skewness spectrum 
of the point sources (DSFG+radio) is a factor of 4-5 higher compared to the case with point source removal 
and becomes the dominant signal at $\ell> 5000$. Since the point sources and the SZ bispectrum have opposite signs, 
the total skewness spectrum (SZ+point sources) is smaller than the SZ skewness spectrum alone, 
especially at higher-$\ell$ where the point source bispectrum becomes non-negligible. 

\subsection{Signal-to-Noise}

Finally, using Eq.~\ref{eq:Bl}, we define the signal-to-noise (S/N) in terms of the 
full-sky bispectrum. The S/N integrated to a certain $\ell$ \citep{ks00, hu00}, 
$\lambda (<\ell)$ or $\lambda_{\ell}$ can be written as 
\be
\lambda(< \ell)=\sqrt {\sum_{\ell_1 \ell_2 \ell_3} \frac{B^2(\ell_1 \ell_2 \ell_3)}{N^2(\ell_1 \ell_2 \ell_3)}}, 
\label{eq:norm}
\ee
where $l_3 \leq l_2 \leq l_1 \leq l$. 
The right panel of Figure~\ref{fig:szbispec} shows the total S/N for the SZ skewness 
spectrum.  For a 2500~$\deg^2$ SPT like survey at 150~GHz, the integrated S/N for the 
SZ skewness spectrum is $\approx 16$. Note that the total (SZ+point source) 
skewness spectrum is lower than the SZ skewness spectrum alone, because these two spectra
have opposite sign.  We find the S/N of the total skewness spectrum is only 
10\% smaller than the SZ only case.  

\section{Summary and Discussions} 

We investigated the SZ bispectrum and the prospects of its detection with the current and 
upcoming CMB surveys. We define a skewness spectrum of the SZ effect which is a sum 
over all possible triangle configurations over the two smaller sides and expressed as a function 
of the largest side. About half of the SZ skewness spectrum signal arises from massive 
($M_{500} \gtrsim 6.5\times 10^{14} h^{-1}M_\odot$) clusters in the local universe 
($z \lesssim 0.4$). 
This is in contrast to the SZ power spectrum whose signals come primarily from high-z, 
low-mass galaxy groups ($M_{500} < 2\times 10^{14} h^{-1}M_{\odot}$ and $z \gtrsim 
0.6$) that have little observational constraints at present. The skewness spectrum, therefore, 
is less susceptible to the large astrophysical uncertainties associated with high-z, low-mass
groups. 

Our model for the SZ profile from groups and clusters reproduces the pressure profile of massive 
X-ray clusters in the local universe. This provides significant constraints on the ICM models of the 
present-day massive clusters leaving only the redshift dependent parameters unknown. Adopting 
wide priors on these parameters, we determine astrophysical uncertainties in the 
skewness spectrum to be about 33\%. This is in contrast to the SZ power spectrum  which is 
uncertain at the level of $50$\%, primarily because a significant fraction of the power comes 
from high-z, low-mass groups with $M_{500} \lesssim 2\times 10^{14} h^{-1}M_\odot$ and 
$z \gtrsim 0.6$. Such objects have not been studied in as much detail as their 
high mass counterparts, making their theoretical modeling and hence the SZ power spectrum highly uncertain. Other than the gas physics uncertainties, the theoretical modeling of the cluster mass function also add to the uncertainties in the bispectrum prediction. Because of the exponential nature of the mass function at the massive cluster end, a small change in cluster masses cause  significant difference in the abundance of massive clusters  \citep{bh11}.

The relation of the amplitude of the SZ skewness spectrum and the SZ power spectrum is robust 
to these astrophysical uncertainties. Indeed, the overall uncertainty of the relation is only 7\% 
within the allowed range of ICM parameters. 
The skewness and the power spectrum amplitudes of the thermal SZ effect is related as 
$B^{\Lambda}_{\rm SZ}=A_{\rm tSZ}^{1.4}$; consequently, the skewness spectrum varies as
$B^{\Lambda}_{\rm SZ }\propto \sigma_8^{11.6}$. Given the astrophysical uncertainty of 33\%, 
the skewness spectrum measurements have the potential to constrain $\sigma_8$ to better than 
3\% accuracy. Since the SZ skewness spectrum gets its contribution primarily from the thermal 
component, the SZ skewness spectrum measurements can constrain the thermal SZ component 
uniquely. We can then use the skewness spectrum measurements to break the degeneracy
between thermal and the kinetic power spectrum measurements.

We show that the current high resolution experiments, such as ACT, Planck, and SPT, have the 
potential to measure the skewness spectrum. For example, using the full 2500 deg$^2$ of data, 
SPT should be able to detect the SZ skewness spectrum with a signal-to-noise $\approx$ 16. 
Indeed, the SZ skewness signal we predict in this study is detected by the ACT collaboration 
 at 5$\sigma$ \citep{act_skewness12}.  
We note however that our noise calculation does not account for all possible sources, 
as we make several simplified assumptions. For example, the contribution of the trispectrum is
assumed to be negligible and we ignore any spatial correlation between the tSZ and CIB.  
However, as most of the SZ trispectrum signal comes from massive clusters \citep{shaw09}, 
masking out these objects should help reduce the SZ trispectrum contribution to the non-Gaussian component of the variance of 
the skewness spectrum. Also we assume the lensing 3-pt function and its correlation with the 
SZ 3-pt function is zero, because CMB lensing dominates at $\ell\le 2000$ while the SZ skewness
spectrum gets most of the signal at $\ell\sim 4000$ (see Fig~8, right panel). In this study we also assume the kSZ bispectrum 
is negligible. While this assumption is valid for the linear part of the kSZ effect, the non-linear kSZ 
effect contributes a non-zero bispectrum which is not included in our calculation. However, 
given that the overall magnitude of the kSZ effect is about 20-40\% of the thermal SZ amplitude, 
the non-linear kSZ bispectrum should be small. Nevertheless, in order to use the SZ skewness
spectrum as a robust cosmological probe, these assumptions must be tested using mock catalogs. 
We will present such study in our future work.

\acknowledgements
We acknowledge useful discussions with John Carlstrom, Salman Habib, and Katrin Heitmann. We also thank the anonymous referee for useful suggestions. 
As this work neared completion, we learned about similar work by members of the ACT
collaboration with whom we exchanged correspondence.
This work was supported by the NSF grant AST-1009811. SB also acknowledges support by
the NSF grant ANT-0638937 and by Argonne National Laboratory's resources under U.S.
Department of Energy contract DE-AC02-06CH11357.  DN acknowledges support by NASA
ATP grant NNX11AE07G, NASA Chandra Theory grant GO213004B, Research Corporation, 
and by Yale University. TC is supported by the NSF grant ANT-0638937. GH is supported by 
NSERC and CIfAR and thanks KICP and FNAL for their hospitality.

\bibliographystyle{apj}

\end{document}